\documentclass{aa}
\usepackage[varg]{txfonts}
\usepackage{amssymb}
\usepackage{multirow}
\usepackage{natbib}
\usepackage{float}
\usepackage{threeparttable}
\usepackage{booktabs}
\usepackage{siunitx}
\usepackage{hyperref}

\hypersetup{
    colorlinks=true,
    linkcolor=blue,
    filecolor=magenta,      
    urlcolor=cyan,
    citecolor=blue,
    }

\bibpunct{(}{)}{;}{a}{}{,}

\begin{document}

\title{Achondrites in meteor data: Spectra, dynamics, and physical properties of candidate aubrite and eucrite impactors}
\titlerunning{Achondrites in meteor data}
\authorrunning{Matlovič et al.}

\author{P. Matlovič\inst{1}
  \and A. Pisarčíková\inst{2}
  \and V. Pazderová\inst{1}
  \and T. Vörös\inst{1}
  \and F. Hlobik\inst{1}
  \and H.A.R. Devillepoix\inst{3,4}
  \and J. Borovička\inst{2}
  \and M. Paprskárová\inst{1}
  \and S. E. Deam\inst{3,4}
  \and J. Tóth\inst{1}
  \and L. Kornoš\inst{1}
  \and T. Paulech\inst{1}
  \and P. Zigo\inst{1}
  }

\institute{Faculty of Mathematics, Physics and Informatics,
  Comenius University Bratislava, Mlynská dolina, 84248 Bratislava, Slovakia\\
  \email{matlovic@fmph.uniba.sk}  
  \and Astronomical Institute of the Czech Academy of Sciences, Fričova 298, 25165 Ondřejov, Czech Republic
  \and Space Science and Technology Centre, School of Earth and Planetary Sciences, Curtin University, Perth, WA 6845, Australia
  \and International Centre for Radio Astronomy Research, Curtin University, Perth WA 6845, Australia
  }

\date{Accepted for publication 01/2026}

\abstract{Meteor spectroscopy presents new opportunities for investigating the diversity of small Solar System bodies and capturing the real distribution of present material types. In this work we analyzed a sample of 180 higher-resolution meteor spectra from the All-sky Meteor Orbit System (AMOS) network to search for meteoroids with atypical compositions. In addition to several iron bodies, we have identified the first two achondritic meteoroids in our database, both likely meteorite-dropping impactors. We analyzed the two cases in detail using their spectral, dynamical, and physical properties, and compared them with a reference ordinary chondrite meteoroid observed under similar conditions. The spectral analysis revealed atypical features in the two achondrites -- strong Mg and Si and low Fe in one case, and strong Ca, Al, and Ti and low Mg in the other. The measured relative elemental abundances imply an aubrite- and a eucrite-like composition. The aubrite-like meteoroid showed an unexpected enhancement in Ca, Mn, and Ti with short-lived intensity spikes not seen in the eucrite-like case, which we interpret as the rapid release of localized inclusions rather than a bulk enrichment. This indicates that transient spectral features can reveal internal heterogeneity in achondritic meteoroids beyond their average composition. The classification of both meteoroids was found to be consistent with the determined dynamical and physical properties. The eucrite meteoroid originated from an orbit affected by the $\nu_6$ resonance in the inner main belt, a common delivery mechanism of Howardite-Eucrite-Diogenite meteorites, and exhibited ablation behavior corresponding to a compact material with low erosion and an estimated bulk density of $\approx$ 3.16 $\pm$ 0.10 g\,cm\textsuperscript{-3}. The aubrite meteoroid originated from a short-period, low-eccentricity orbit similar to some known E-type near-Earth asteroids. Both events also exhibited atypical light curve behavior, but our results indicate that the robust identification of achondritic meteoroids in meteor surveys generally requires emission spectra. This work presents one of the first detailed studies of achondritic meteoroids from meteor observations and aims to provide reference properties of atypical meteors for more efficient identifications of achondrites in future surveys.}

\maketitle

\section{Introduction} \label{sec:intro}

Advancements in the coverage and fidelity of systematic meteor observations offer new opportunities for identifying impactors with atypical compositions. Most information on achondritic materials comes from laboratory meteorite studies, which can be linked with reflectance spectra of asteroid surfaces to identify their potential parent objects. The classification of meteoroid composition types in larger meteor datasets can help describe the sources of these bodies and their behavior during atmospheric entry, and ultimately to characterize the actual distribution of interplanetary materials within the Solar System. This carries important implications for the evolution and migration models of our planetary system.

The identification of atypical meteoroid types from meteor observations has been rare thus far, mainly due to the limited scope and resolution of spectroscopic observations performed globally in the past. The European Fireball Network team has provided early insights into the spectral properties of meteors and fireballs with assumed chondritic-like compositions \citep{1993A&A...279..627B, 2005EM&P...97..279B}. The spectral type of these chondritic-like meteors is often referred to as ``normal,'' based on assumed typical Solar System abundances, and may be consistent with ordinary and carbonaceous chondrites, as well as cometary bodies. Using larger meteor spectral datasets of fainter meteors, meteoroids with spectral features deviating from the assumed chondritic properties, including irons and thermally altered sodium-depleted meteoroids, have been identified \citep{2005Icar..174...15B}. Such bodies were later confirmed by other authors \citep{2019A&A...621A..68V, 2019A&A...629A..71M, 2020P&SS..19405040A}. 

Apart from the easily detectable irons, and one fireball with a spectrum consistent with a diogenite meteoroid reported by \citet{1994ASPC...63..186B, 1994Metic..29..446B}, no achondrites have been identified from meteor observations. Recent large-scale meteorite ablation experiments have provided detailed insights into expected spectral signatures of individual meteorite types under meteor-entry conditions \citep{2024A&A...689A.323M, 2024Icar..40715791T}. Combined with the advancements in meteor spectral data resolution and coverage, this introduces better opportunities to identify rare meteoroids, including achondrites, in meteor datasets. Beyond its contribution to our understanding of the sources and distribution of interplanetary materials, this information is also crucial for prioritizing meteorite recovery searches. This is exemplified by the recent rare, instrumentally observed meteorite fall of the aubrite Ribbeck \citep{2024A&A...686A..67S}, whose atypical composition was predicted from its fireball spectrum.

Here we report the identification of two achondrite meteoroids from our All-sky Meteor Orbit System (AMOS) network database. We characterize in detail their emission spectra, which we measured and modeled to estimate their composition (Sect. \ref{sec:spectral}), their physical properties from ablation and fragmentation modeling (Sect. \ref{sec:physical}), and their dynamical properties, including their orbital evolution (Sect. \ref{sec:dynamical}). This analysis aims to provide a reference for more efficient identifications of similar atypical meteoroids in other datasets.

\section{Instrumentation, data, and methods} \label{sec:methods}

\begin{figure}[]
\centerline{\includegraphics[width=\columnwidth,angle=0]{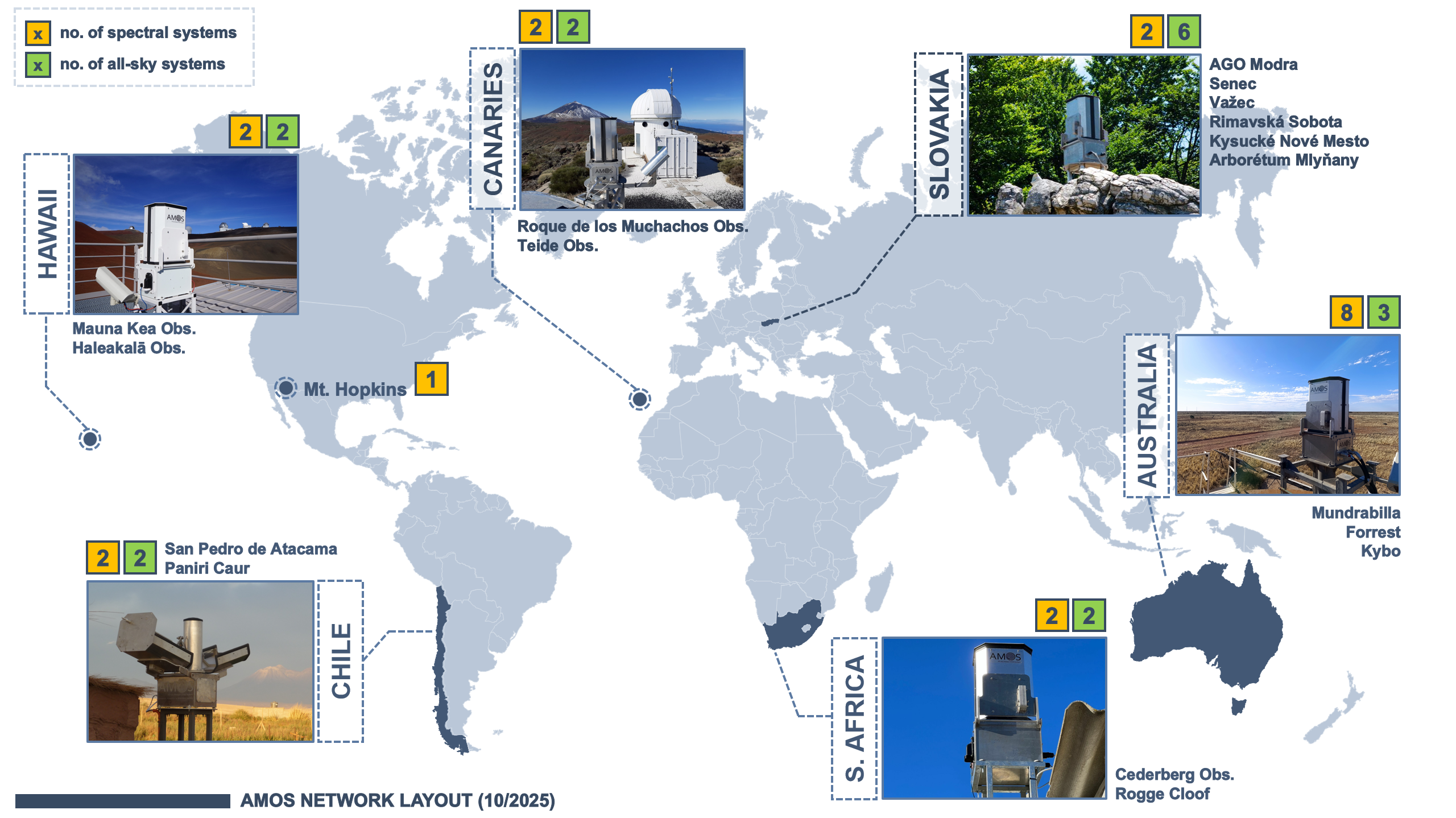}}

\caption[f1]{Illustrative map of the AMOS network station locations, indicating the number of all-sky and spectral systems at each site.}
\label{AMOS_map}
\end{figure} 

The search for meteoroids of atypical composition was conducted using only good-quality (S/N > 10) higher-resolution (FWHM $\le$ 1.3 nm) meteor spectra captured simultaneously by the AMOS spectrographs and multi-station AMOS all-sky stations. The global AMOS network currently consists of 17 all-sky systems and 19 spectrographs at stations in Slovakia, Canary Islands, Chile, Hawaii, Australia, South Africa, and Arizona (Fig. \ref{AMOS_map}).

The AMOS-Spec slitless spectrograph uses a 1000\,gr/mm holographic grating in front of a 6\,mm/f1.4 lens and a DMK~33UX252 digital camera with an 8-bit output, 2048$\times$1536 pixel resolution, operating at a frame rate of 20\,fps. The field of view (FOV) is 60$^{\circ}$$\times$45$^{\circ}$, and the limiting stellar magnitude is +4. The typical achieved spectral resolution is around 1.1--1.3\,nm (FWHM). The spectral reduction and analysis were performed following the procedures described in \citet{2019A&A...629A..71M} and \citet{2022MNRAS.513.3982M}. Each spectrum was manually scanned in individual frames of the video recording, calibrated, and fitted with a synthetic spectrum comprised of the main atomic and molecular emission features of the low- and high-temperature components in meteors. The synthetic spectrum was fitted to the calibrated meteor spectrum using the damped least-squares method (the Levenberg–Marquardt algorithm) within the \textsc{Fityk} software \citep{Wojdyr:ko5121}. Gaussian instrumental line profiles were assumed in the synthetic spectra, and line intensities were measured by integrating the line profiles. The uncertainties of the measured line intensities were calculated from their S/N in each spectrum. The expressed multiplet numbers are taken from \citet{1945CoPri..20....1M}. In addition to standard line intensity measurements, the spectra were fitted using a radiative transfer model from  \citet{1993A&A...279..627B}, which assumes local thermal equilibrium (LTE) and self-absorption within the radiating plasma.

\begin{figure}[]
\centerline{\includegraphics[width=.98\columnwidth,angle=0]{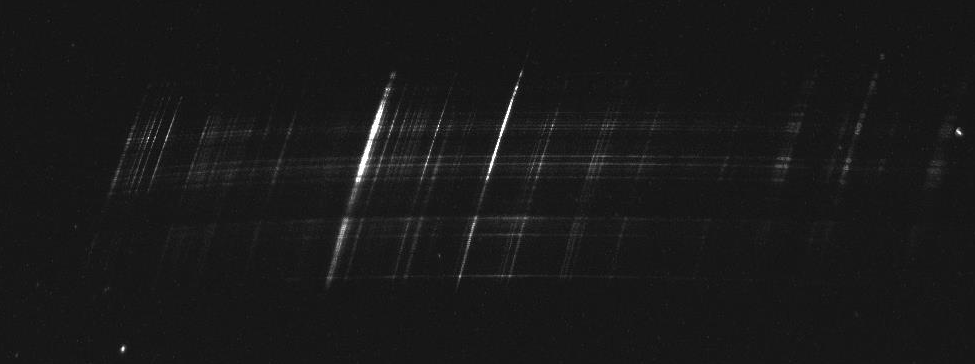}}
\centerline{\includegraphics[width=.98\columnwidth,angle=0]{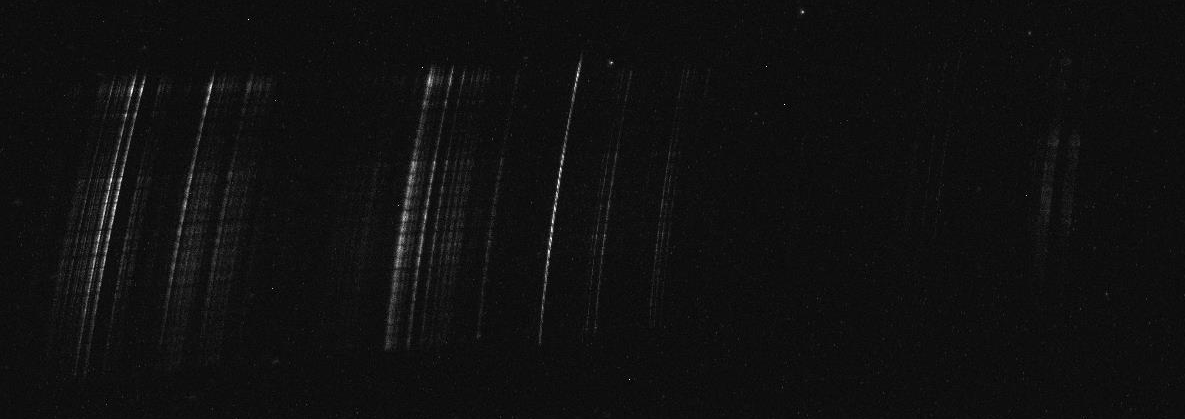}}
\centerline{\includegraphics[width=.98\columnwidth,angle=0]{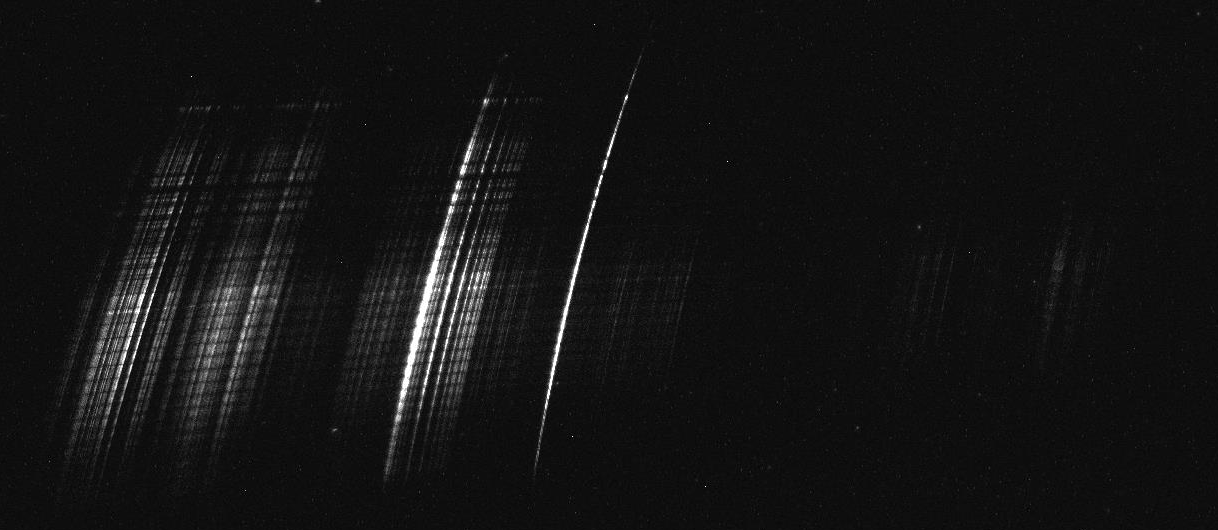}}
\caption[f1]{Stacked images from individual video frames of the spectra of achondrites ACH1 (upper panel), ACH2 (middle panel), and an ordinary chondritic meteoroid, {M20240807\_093858} (lower panel) as captured by the AMOS-Spec systems in Australia (ACH1) and Hawaii (ACH2 and the ordinary chondrite).}
\label{spectra_images}
\end{figure} 

The meteor trajectories, orbits, and photometry were studied based on multi-station observations from all-sky AMOS systems. The instruments consist of a fisheye lens, an image intensifier, a projection lens, and a digital video camera operating at 1600$\times$1200 pixel resolution. The resulting FOV of AMOS is 180$^{\circ}$$\times$140$^{\circ}$ with videos recorded at 20\,fps. This translates to a resolution of 6.8 arcmin pixel$^{-1}$. The limiting magnitude for stars is around +5 mag for a single frame, and approximately +4 mag for objects at typical meteor speeds due to the trailing losses. The meteor trajectories and orbits were measured using the methods described in detail in \citet{AMOSinprep}.

The astrometry and heliocentric orbit determination were performed using our software, Meteor Trajectory (\texttt{MT}), which has been upgraded from the earlier version described in \citet{KornosIMC17}. The astrometric reduction is based on the all-sky procedure from \citet{1995A&AS..112..173B}. The program uses geometric transformations to compute the rectangular coordinates of catalog stars, which are compared with measured stars. From this, 13 plate constants are determined; these yield the zenith distance and azimuth of a meteor. As a reference, the SKY2000 Master Catalog, Version 4 \citep{2001yCat.5109....0M} is used for stars up to the +6th magnitude and includes color indices. The precision of the AMOS astrometry (standard deviation of star positions) is approximately 0.02$^\circ$--0.03$^\circ$. This translates to an accuracy of 10--100\,m for the atmospheric meteor trajectory. The orbit determination method within \textsc{\texttt{MT}} is based on the plane intersection method of \citet{1987BAICz..38..222C}, with custom modules for the speed-fit models, time-shift analysis, and Monte Carlo simulations for error estimation. The meteor photometry is implemented in the same software. It uses a brightness calibration based on reference stars for meteors $\geq$\,$-2$\,mag, and a saturation correction for brighter fireballs, based on a comparison with bright planets and the Moon in different phases. Each meteor in this work was processed manually.

The dynamical evolution of studied meteoroids was investigated via backward orbital integrations using the IAS15 integrator included in the REBOUND software package \citep{2012A&A...537A.128R}. IAS15 is a non-symplectic 15th-order integrator with adaptive step-size control \citep{2015MNRAS.446.1424R}. The integrations take into account the gravitational effects of the Sun, planets, the Moon, and the four largest asteroids from the main belt: Ceres, Pallas, Vesta, and Hygiea. Nongravitational effects were not modeled. 

Meteor trajectory observations for one of the achondrites were provided by the Desert Fireball Network (DFN; \citealt{2019MNRAS.483.5166D}). DFN and AMOS collaborate in the Nullarbor Plain region in Australia to cover dynamical and spectral observations of both fainter meteors and fireballs \citep{2022RNAAS...6..144D}. The trajectory and orbit for this case were determined using the standard methods of the DFN pipeline described in \citet{2022M&PS...57.1328D}. AMOS and DFN data reduction pipelines were recently described and compared with other networks in \citet{2026A&A...705A..65S}.

\section{Spectra and composition} \label{sec:spectral}

\begin{figure*}[]
\centerline{\includegraphics[width=.85\textwidth,angle=0]{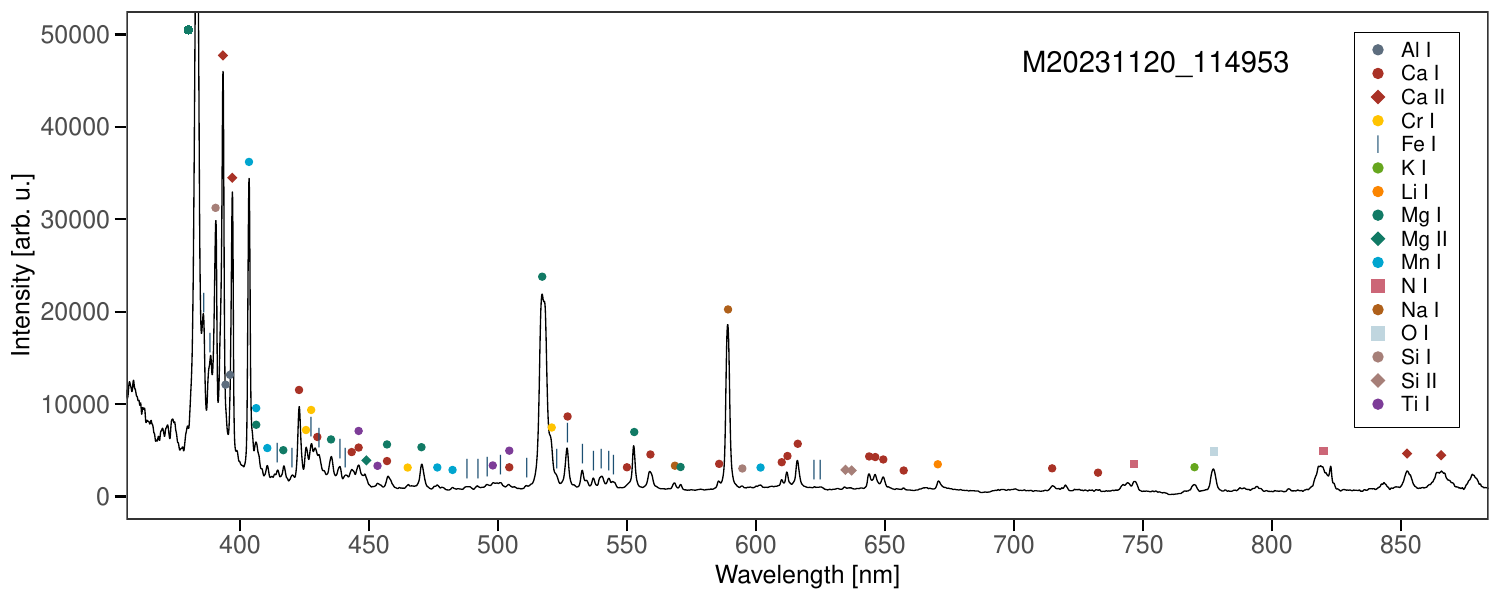}}
\centerline{\includegraphics[width=.85\textwidth,angle=0]{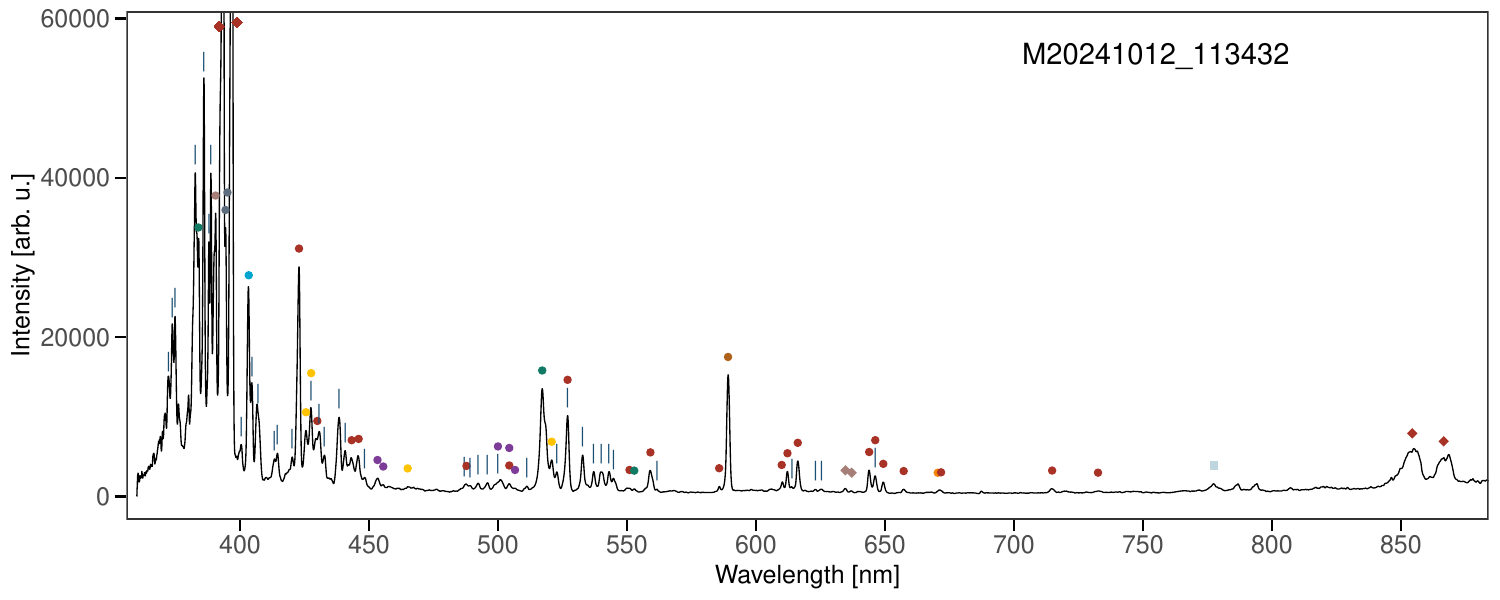}}
\caption[f1]{Full calibrated spectral profiles of meteors ACH1 (the aubrite candidate) and ACH2 (the eucrite candidate). The most important identified emission lines are labeled.}
\label{spectra_profiles}
\end{figure*} 

\begin{figure*}[]
\centerline{\includegraphics[width=.75\textwidth,angle=0]{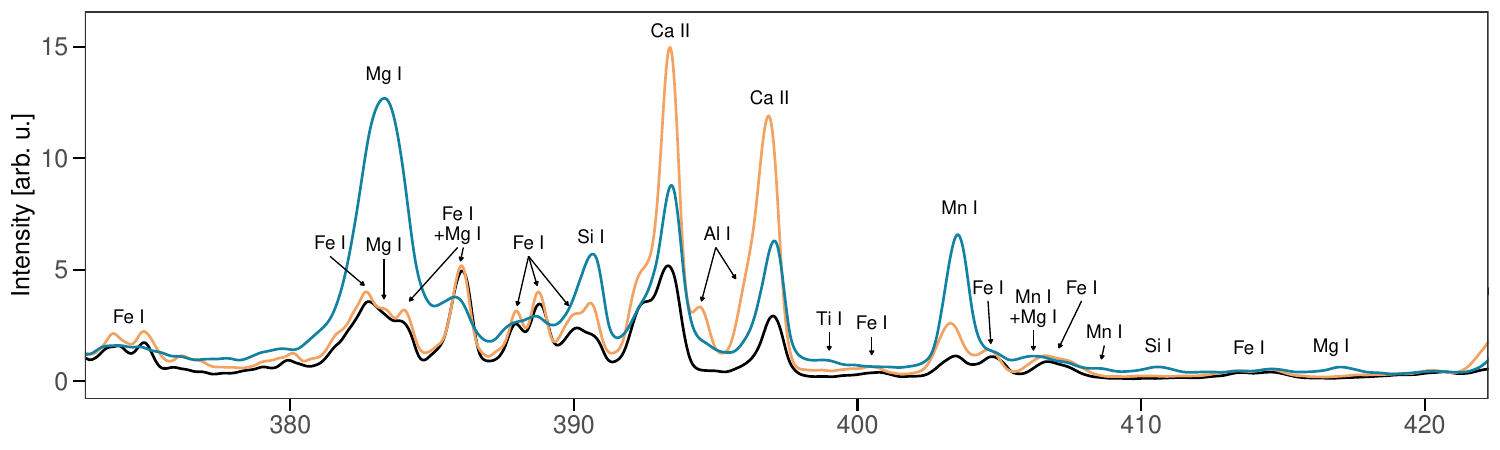}}
\centerline{\includegraphics[width=.75\textwidth,angle=0]{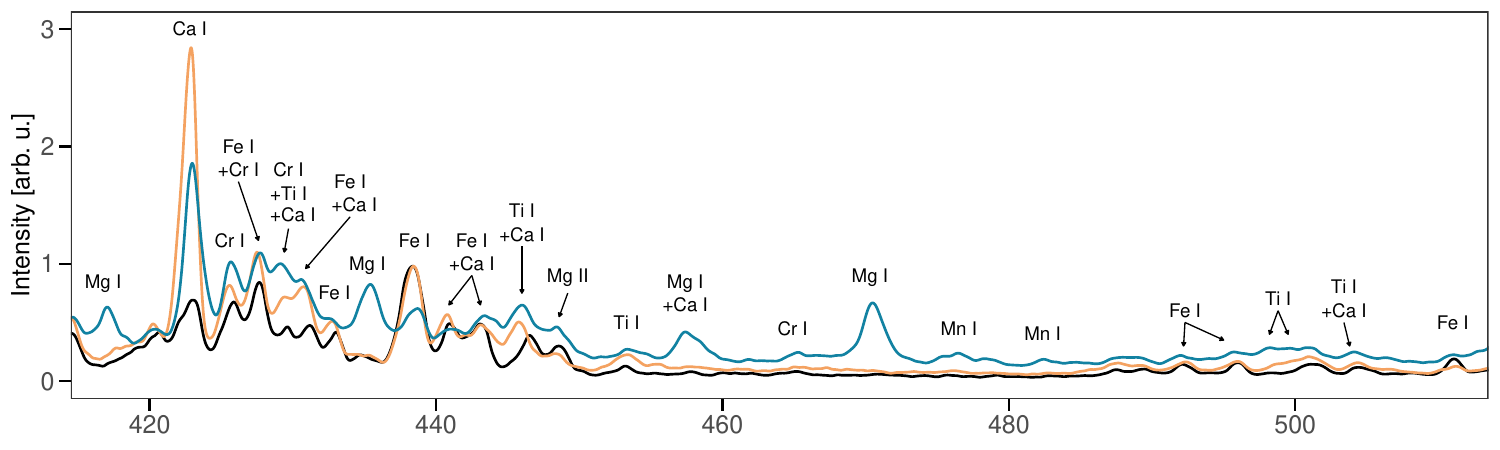}}
\centerline{\includegraphics[width=.75\textwidth,angle=0]{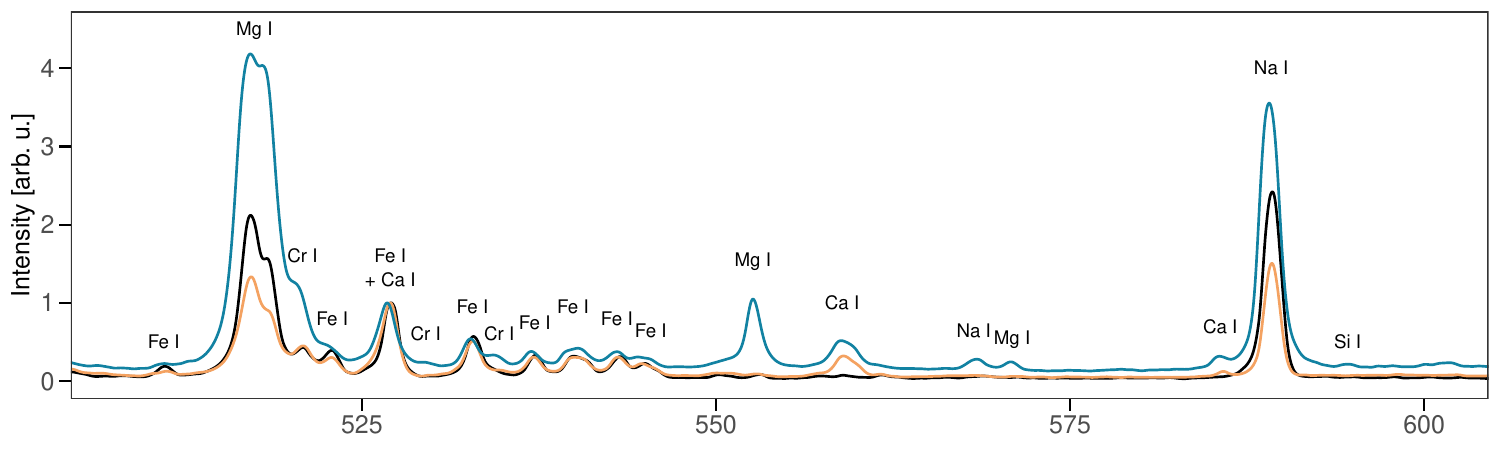}}
\centerline{\includegraphics[width=.75\textwidth,angle=0]{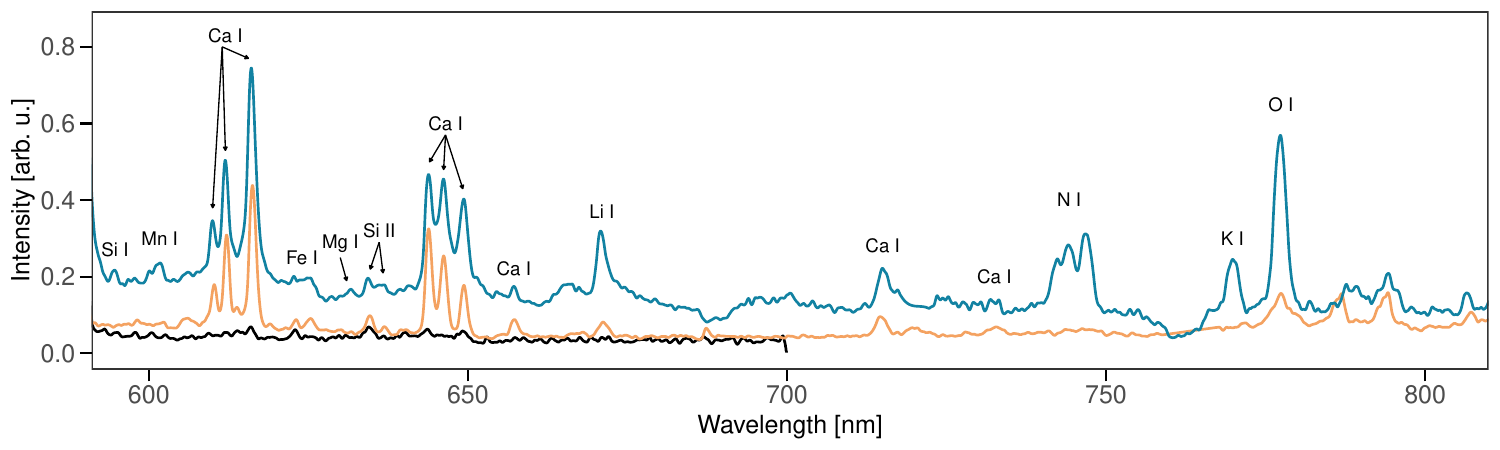}}
\caption[f1]{Comparison of the spectra of the two achondrites --- the aubrite candidate (ACH1; in blue) and the eucrite candidate (ACH2; in orange) --- with a typical ordinary-chondrite-like spectrum (black). The plots are split into separate wavelength ranges for better visibility, with intensities normalized to unity at the peak of the Fe I line at $\approx$ 527 nm. The aubrite candidate spectrum includes a more notable continuum and faint molecular bands of N\textsubscript{2}, which can be seen in the bottom two panels.}
\label{spectra_comparison}
\end{figure*}

While the ablation behavior of the two studied meteoroids (further discussed in Sect. \ref{sec:physical}) provides some indication of atypical material properties, the identification of the achondritic composition of these meteoroids was only possible by analyzing their emission spectra. Images of the recorded spectra of the two achondrites are shown in Fig. \ref{spectra_images}, along with an image of a common chondritic asteroidal meteoroid spectrum for comparison. Spectral profiles of the two achondrites with all identified emission lines are displayed in Fig. \ref{spectra_profiles}. The spectra show notable differences from typical chondritic and cometary meteors. A detailed comparison of the spectra of the two achondrites with a typical asteroidal chondritic meteor spectrum is given in Fig. \ref{spectra_comparison}. 

The spectrum of the first case -- meteor {M20231120\_114953} (hereafter ACH1) -- shows distinctly strong intensities of Mg, Si, Mn, Ca, Ti, and Li, and a low Fe intensity. Among our entire AMOS archive of more than 500 processed meteor spectra, we identify this as the case with the strongest relative Mg, Si, Mn, and Li emission, based on the measured line intensities normalized to the total recorded spectral intensity. These features are consistent with an achondritic composition enriched in enstatite. The spectrum also shows strong features of the refractory elements Ca and Ti, with intensities comparable only to those of the other achondrite, {M20241012\_113432} (hereafter ACH2). In contrast to ACH2, ACH1 shows fainter Ca I and Ca II lines, and less prominent or absent Al I lines. The measured line intensities of some refractory species can be partly affected by the lower speed of the ACH1 meteor (13.9\,km\,s$^{-1}$). Lower entry speeds are generally associated with lower ablation temperatures. The incomplete evaporation of refractory elements in meteors has already been noted by \citet{1993A&A...279..627B} and was recently confirmed to be pronounced for conditions comparable to very low-speed meteors simulated in plasma wind tunnel experiments, where no Ca, Al, or Ti emission was detected, even for refractory-rich meteorites \citep{2024A&A...689A.323M}. Since relatively strong Ca and Ti lines have been detected in ACH1, the lack of notable Al I emission near 395\,nm likely reflects a composition not significantly enhanced in Al.

Based on the analyzed spectral line intensities, and considering the typical bulk composition of various meteorite types and their spectral signatures discussed in \citet{2024A&A...689A.323M}, we infer that the ACH1 spectrum is most consistent with the composition of an aubrite. The spectrum is very similar to the recent instrumentally observed meteorite fall -- the aubrite Ribbeck \citep{2024A&A...686A..67S} -- including its Mg-rich, Fe-poor features and enhanced intensities of Ca, Ti, Li, and K. In contrast, our case exhibits fainter or absent Al, slightly weaker Na, and stronger Mn emission. 

The spectrum of the second case, ACH2, shows some similarities to the previous case, but, unlike the enstatite-rich ACH1, this case shows notably weaker Mg emission and stronger Fe lines (Fig. \ref{spectra_profiles}). The most distinct features of the ACH2 spectrum are the Ca I and Ca II lines, which are even stronger than in ACH1, as well as higher intensities of Ti I, Al I, Mn I, Cr I, and Li I. This case shows the strongest relative Ca emission among all analyzed AMOS spectra, based on the Ca I and Ca II line intensities expressed as a fraction of the total recorded spectral intensity. Furthermore, it exhibits an atypically high Si I/Mg I line intensity ratio that has not been observed in any other spectrum in our archive. The observed spectral properties are consistent with a meteoroid composition similar to eucrites or howardites. This is also supported by the measured Fe/Mg and Cr/Mg line intensity ratios, which are higher than those of typical chondritic spectra (Fig. \ref{spectra_comparison}), in line with our findings from ablated Howardite-Eucrite-Diogenite (HED) meteorite spectra \citep{2024A&A...689A.323M}.

To examine evidence of differential ablation within the achondritic meteors, we additionally analyzed the monochromatic light curves of both events, which show time-resolved emission of individual species (Mg I, Fe I, Na I, Ca I, Mn I, and Li I). For the spectrally recorded bright part of the ACH2 fireball, no systematic differences consistent with differential ablation were detected, with all studied species following the same light curve patterns. While differential ablation was suspected for this eucrite candidate based on its overall light curve, which shows a notable transition phase (Sect. \ref{sec:physical}), this interval was not covered by spectral data due to insufficient brightness. The monochromatic light curves of ACH1 show numerous flares, which are for the most part consistent for the different species and likely reflect individual fragmentation events. However, for some species, the relative intensities deviate from the overall trend at different times, which we interpret as the preferential ablation of compositionally distinct phases (discussed later in this section).

While the measured line intensity ratios already provide clues as to the composition of the two meteoroids, we aimed to confirm this inference by measuring relative abundances of main chemical elements in the observed meteor plasma using a spectrum model introduced by \citet{1993A&A...279..627B}. This radiative transfer model assumes LTE in the radiating meteor plasma and takes self-absorption in the optically thick environment into account. First, the plasma conditions, including the temperature and column density, were derived by fitting numerous Fe I lines. Then, relative abundances of other species with respect to Fe were determined via a careful fitting of the full spectrum. For ACH1 and ACH2, the fitting was performed on the brightest part of the spectrum (approximately ten frames) to ensure roughly uniform plasma conditions. For a representative ordinary chondrite, a summed spectrum was used instead, constructed from the highest-quality frames, as the brightest part was significantly saturated. 

The radiating plasma parameters obtained from spectral fitting, along with relative abundances of atoms before ionization correction, are listed in Table \ref{tab:plasma_conditions}. Lines deviating from the LTE assumption or affected by saturation were omitted from the fit. The abundances of neutral atoms were corrected for ionization to obtain the true relative elemental abundances. The ionization correction was determined using the Saha equation following the method of \citet{1993A&A...279..627B}, for which we used the meteor’s known brightness to estimate the electron density. The brightness defines the cross section (Table \ref{tab:plasma_conditions}). Assuming a 2:1:1 shape for the radiating volume and knowing the geometry of observation, the line-of-sight dimension could be computed and the column densities of neutral atoms could be converted to volume densities. Considering the contributions of electrons from all elements, the total electron density could then be computed (see \citealt{1993A&A...279..627B} for details). The resulting electron densities are \num{4.3e14}cm\textsuperscript{-3}, \num{2.3e12}cm\textsuperscript{-3}, and \num{4.4e12}cm\textsuperscript{-3} for ACH1, ACH2, and ordinary chondrite, respectively. The resulting relative abundances of the two achondritic meteoroids compared to a representative chondritic asteroidal meteoroid are summarized in Table \ref{tab:abundances}. The abundance uncertainties were determined by estimating upper and lower abundance limits through a comparison of the fitted and observed spectra in the regions around the relevant lines, combined with the variation in the abundance values caused by changes in the plasma conditions.

\begin{table}[h]
\centering
\caption{Radiative-transfer fit parameters and pre-correction relative abundances for ACH1, ACH2, and a reference ordinary chondrite.}
\label{tab:plasma_conditions}
\resizebox{\columnwidth}{!}{%
\begin{tabular}{lccc}
\toprule
\multicolumn{1}{c}{Meteor} & M20241012\_113432 & M20231120\_114953 & M20240807\_093858\\
& (ACH2, eucrite) & (ACH1, aubrite) & (ordinary chondrite)\\
\midrule
T {[}K{]} & 4200 $\pm$ 350 & 4800 $\pm$ 350 & 4200 $\pm$ 250\\
N\textsubscript{Fe I} {[}cm\textsuperscript{-2}{]} & \num{7.6e+14} $\pm$ \num{5.5e+14} & \num{3.1e+15} $\pm$ \num{1.9e+15} & \num{1.1e+15} $\pm$ \num{1.7e+14} \\
$\Gamma$ {[}s\textsuperscript{-1}{]} & \num{1.0e+10} $\pm$ \num{1.0e+9} & \num{1.4e+10} $\pm$ \num{8.0e+9} & \num{5.0e+09} $\pm$ \num{2.4e+8} \\
P {[}m\textsuperscript{2}{]} & 200 $\pm$ 170 & 3 $\pm$ 7 & 20 $\pm$ 15 \\
\midrule
Si I/ Fe I & 23.00 $\pm$ 4.50 & 80.00 $\pm$ 10.00 & 3.00 $\pm$ 0.75 \\
Mg I/ Fe I & 0.80 $\pm$ 0.09 & 260.00 $\pm$ 20.00  & 1.83 $\pm$ 0.20 \\
Ca I/ Fe I & 0.011 $\pm$ 0.002 & 0.033 $\pm$ 0.006 & 0.0002 $\pm$ 0.00003 \\
Al I/ Fe I & 0.10 $\pm$ 0.04 & 0.04 $\pm$ 0.03 & 0.001 $\pm$ 0.0006 \\
Ti I/ Fe I & 0.001 $\pm$ 0.0001 & 0.003 $\pm$ 0.001 & 0.0004 $\pm$ 0.00005 \\
Cr I/ Fe I & 0.004 $\pm$ 0.0003 & 0.048 $\pm$ 0.007 & 0.002 $\pm$ 0.0004 \\
Mn I/ Fe I & 0.02 $\pm$ 0.002 & 0.095 $\pm$ 0.033 & 0.006 $\pm$ 0.0007\\
Na I/ Fe I & 0.0002 $\pm$ 0.00001 & 0.032 $\pm$ 0.006 & 0.002 $\pm$ 0.001 \\
K I/ Fe I & \num{1e-6} $\pm$ \num{3e-7} & 0.0002 $\pm$ 0.00005 & - \\
Li I/ Fe I & \num{1e-5} $\pm$ \num{2e-6} & \num{6e-5} $\pm$ \num{2e-6} & -\\
\bottomrule
\end{tabular}}
\tablefoot{Results of spectral fitting for ACH1 (aubrite candidate), ACH2 (eucrite candidate), and a representative ordinary chondrite. Listed are the parameters of the main spectral component derived from the radiative transfer model: temperature ($T)$, column density of Fe\,I atoms $(N_{\mathrm{Fe\,I}}$), damping constant $(\Gamma$), and radiating plasma cross section $(P$), together with abundances of selected species relative to Fe\,I (before ionization correction).}
\end{table}

\begin{table*}[h]
\centering
\caption{Elemental mass fractions of selected species in the meteor plasma, expressed relative to Mg and Fe, derived from the radiative transfer model for ACH1 (aubrite candidate), ACH2 (eucrite candidate), and a representative ordinary chondrite.}
\label{tab:abundances}
\resizebox{.90\textwidth}{!}{%
\begin{tabular}{lcccccc}
\toprule
& \multicolumn{3}{c}{Relative to Mg} & \multicolumn{3}{c}{Relative to Fe} \\
\cmidrule(lr){2-4}\cmidrule(lr){5-7}
Element &
M20241012\_113432  &
M20231120\_114953  &
M20240807\_093858  &
M20241012\_113432  &
M20231120\_114953  &
M20240807\_093858  \\
 &
(ACH2, eucrite) &
(ACH1, aubrite) &
(ordinary chondrite) &
(ACH2, eucrite) &
(ACH1, aubrite) &
(ordinary chondrite) \\
\midrule
Si & 19.83 $\pm$ 6.23 & 0.34 $\pm$ 0.05 & 1.35 $\pm$ 0.37 & 9.37 $\pm$ 2.09 & 39.36 $\pm$ 5.01 & 1.32 $\pm$ 0.34 \\
Mg & 1.00 $\pm$ 0.00 & 1.00 $\pm$ 0.00 & 1.00  $\pm$ 0.00 & 0.47 $\pm$ 0.07 & 117.2 $\pm$ 9.90 & 0.98 $\pm$ 0.12 \\
Fe & 2.12 $\pm$ 0.26  & 0.01 $\pm$ 0.0003 & 1.02 $\pm$ 0.06 & 1.00 $\pm$ 0.00 & 1.00 $\pm$ 0.00 & 1.00 $\pm$ 0.00 \\
Ca & 0.72 $\pm$ 0.27 & 0.001 $\pm$ 0.0005 & 0.004 $\pm$ 0.001 & 0.34 $\pm$ 0.15 & 0.08 $\pm$ 0.06 & 0.004 $\pm$ 0.001 \\
Al & 0.59 $\pm$ 0.27 & 0.0002 $\pm$ 0.00015 & 0.002 $\pm$ 0.001 & 0.28 $\pm$ 0.14 & 0.02 $\pm$ 0.017 & 0.002 $\pm$ 0.001 \\
Ti & 0.014 $\pm$ 0.004 & 0.00003 $\pm$ 0.00001 & 0.002 $\pm$ 0.0002 & 0.007 $\pm$ 0.002 & 0.003 $\pm$ 0.002 & 0.002 $\pm$ 0.0003\\
Cr & 0.024 $\pm$ 0.006 & 0.0004 $\pm$ 0.0001 & 0.005 $\pm$ 0.001 & 0.011 $\pm$ 0.004 & 0.052 $\pm$ 0.010 & 0.005 $\pm$ 0.001 \\
Mn & 0.068 $\pm$ 0.005 & 0.0008 $\pm$ 0.0003 & 0.008 $\pm$ 0.001 & 0.032 $\pm$ 0.005 & 0.097 $\pm$ 0.033 & 0.008 $\pm$ 0.001 \\
Na & 0.033 $\pm$ 0.011 & 0.0009 $\pm$ 0.0007 & 0.079 $\pm$ 0.038 & 0.016 $\pm$ 0.006 & 0.104 $\pm$ 0.089 & 0.078 $\pm$ 0.037 \\
K  & 0.002 $\pm$ 0.0007 & \num{4e-5} $\pm$ \num{3e-5} & -- & 0.001 $\pm$ 0.0004 & 0.005 $\pm$ 0.004 & -- \\
Li & \num{2e-4} $\pm$ \num{8e-5} & \num{3e-7} $\pm$ \num{2e-7} & -- & \num{1e-4} $\pm$ \num{4e-5} & \num{4e-5} $\pm$ \num{3e-5} & -- \\
\bottomrule
\end{tabular}}
\end{table*}

The determined elemental abundances of the meteor plasma strongly support the aubrite and eucrite composition indicated by the anomalous line intensities. The most important diagnostic features suggesting a eucrite composition for meteoroid ACH2 are the very high Si/Mg ratio ($\approx$\,29 in the radiating plasma) and high concentrations of Ca, Al, and Ti. The high Mn and Li abundances and low Na are also consistent with this interpretation. We note that, according to our data, this meteoroid could also be consistent with a howardite composition. Howardites are formed as regolith breccias of eucrite and diogenites and can have bulk compositions similar to those of eucrites. Both classes are thought to originate from V-type parent asteroids. An alternative, less plausible interpretation of the observed spectrum is a composition similar to angrites or shergottites. Both these classes can contain relatively high Fe content and moderate to low Mg, and are rich in refractory elements and Mn. When compared to meteorite fall statistics on Earth, these meteorite types are expected to be less common in interplanetary space compared to HED-type materials. Furthermore, the dynamical properties of this meteoroid (Sect. \ref{sec:dynamical}) also support an origin from a V-type parent body.

The measured elemental abundances in the ACH1 plasma confirm the Mg-rich and Fe-poor composition of this meteoroid, consistent with aubrites. Since the relative abundances of other species are expressed relative to Mg and Fe (Table \ref{tab:abundances}), their values are affected by the extreme abundances of Mg and Fe. The proposed aubrite meteor plasma shows notably high ratios of Mg/Fe\,$\approx$\,117 and Si/Fe\,$\approx$\,39. 

Interestingly, the ACH1 spectrum also shows relatively strong Ca, Mn, and Ti lines. Bulk compositions of Fe-poor aubrite meteorites are typically not enriched in refractory elements (Ca and Ti) or Mn \citep{2003GeCoA..67..557L}. We interpret the inferred high content of Ca, Mn, and Ti in the aubrite candidate as the result of preferential ablation of accessory highly reduced phases, such as oldhamite (CaS) and related sulfide or metal blebs that concentrate Ca, Mn, Ti, and Cr in localized pockets, rather than reflecting the average silicate (enstatite-dominated) composition of the meteoroid \citep[e.g.,][]{PAPIKE_1998_achondrites, 1989Metic..24..195K}. These sulfide-rich droplets are known  to form immiscible melts in enstatite achondrites during high-temperature processing and impact reheating, and can volatilize rapidly when first exposed to atmospheric entry heating, injecting Ca, Mn, and Ti efficiently into the meteor plasma. In contrast, in more oxidized, plagioclase-bearing meteoroids (such as basaltic achondrites and chondrites), Ca and Ti are primarily bound in refractory silicate lattices (plagioclase, high-Ca pyroxene, and ilmenite), which must first melt and then gradually vaporize, producing a more homogeneous and temporally delayed release of these elements. 

\begin{figure}[]
\centerline{\includegraphics[width=\columnwidth,angle=0]{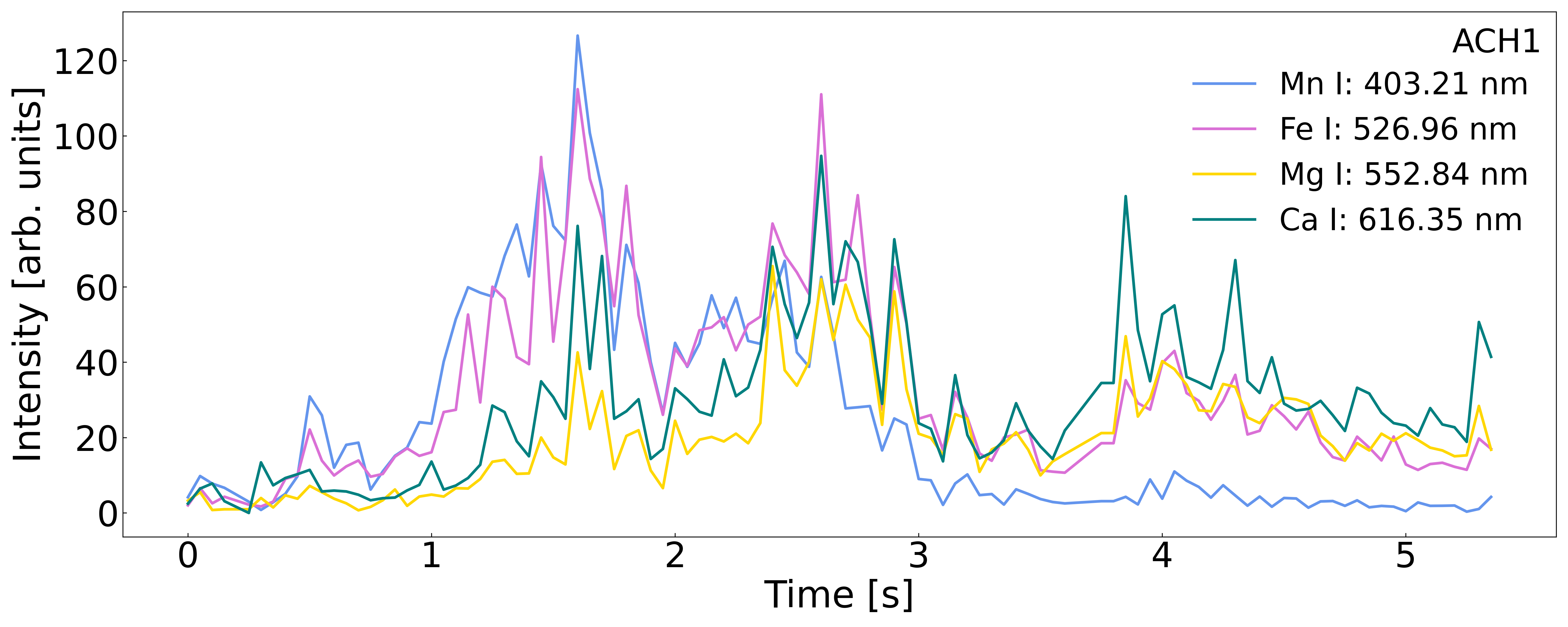}}
\centerline{\includegraphics[width=\columnwidth,angle=0]{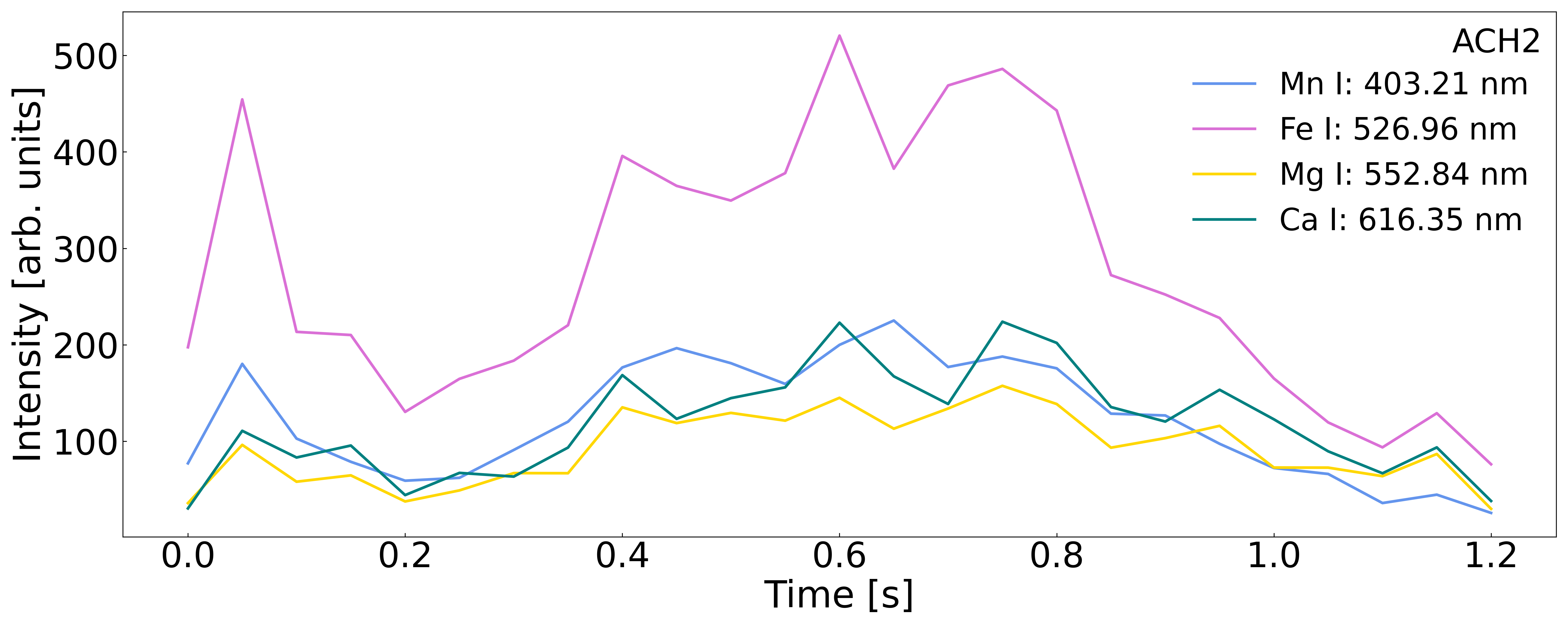}}
\caption[f1]{Monochromatic light curves of the two candidate achondrites: ACH1 (upper panel) and ACH2 (lower panel), comparing the time-resolved emission of selected species, Mg I, Fe I, Mn I, and Ca I, based on selected well-resolved spectral lines. The time on the x-axis begins from the first recorded frame of the meteor spectrum.}
\label{monochroma}
\end{figure} 

\begin{figure}[]
\centerline{\includegraphics[width=.9\columnwidth,angle=0]{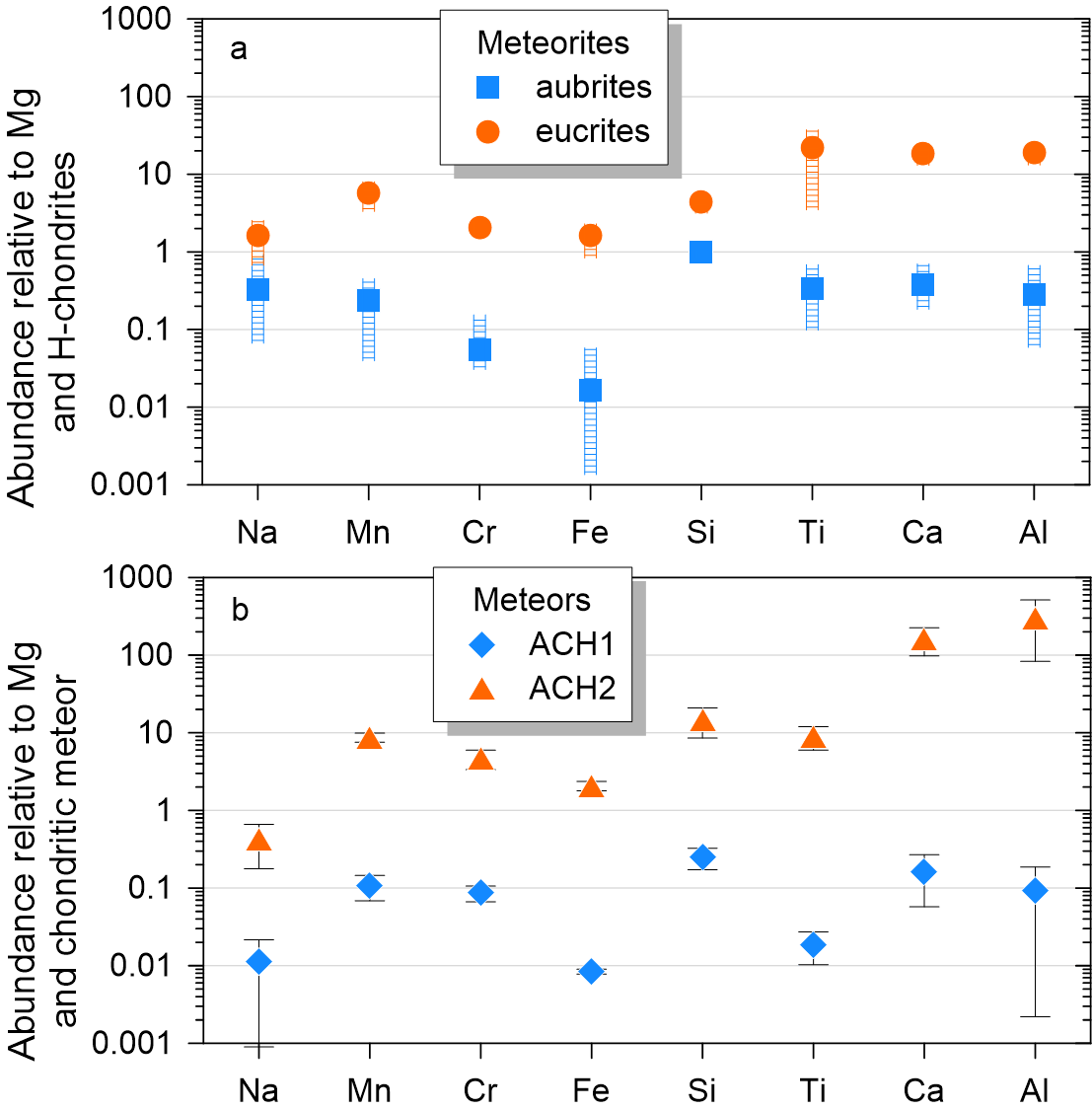}}
\caption[f1]{Upper panel: Abundances of eight elements expressed relative to Mg and H chondrites in aubrites and eucrites. The average value and the full range of values for 13 Fe-poor aubrites are shown \citep{2003GeCoA..67..557L}. The average value and the full range of values for 20 eucrites are shown \citep{1998M&PSA..33..197K}. The average abundances of H chondrites were taken from \citet{1988RSPTA.325..535W}. Lower panel: Relative abundances to Mg and their uncertainties in two achondritic meteors shown in relation to the normal chondritic meteor (Table \ref{tab:abundances}).}
\label{abundances_comp}
\end{figure} 

This interpretation is consistent with the observed distinction in monochromatic light curves between the two achondritic meteors (Fig. \ref{monochroma}). The emission of Ca and Mn in the aubrite candidate is observed in sharp flares, which sometimes deviate in relative brightness from the emission of other species (Fe I and Mg I). The emission of Mn I is low in the second half of the spectral recording, possibly affected by stronger atmospheric absorption in the near-UV in this interval. In contrast, the monochromatic light curves of the eucrite candidate imply a gradual release of these species that is not accompanied by strong flares and follows the same relative intensity pattern throughout the luminous trajectory. Similarly to our aubrite candidate, strong Ca, Ti, and Mn intensities were also detected in the fireball spectrum of the confirmed aubrite meteorite fall Ribbeck \citep{2024A&A...686A..67S}. 

The ordinary chondrite {M20240807\_093858}, analyzed for reference and comparison to asteroidal meteors with more typical compositions, showed higher Fe and Na content than both achondrites, and higher Mg content than the eucrite. The relative abundances found for this case are consistent with the assumed ordinary chondrite composition. While some of the aforementioned elemental ratios found in achondritic plasma may appear extreme compared to average bulk composition ratios in achondritic meteorites, it must be noted that the composition of the radiating meteor plasma is in most cases not fully representative of the original meteoroid composition, due to the effects of incomplete evaporation \citep{1993A&A...279..627B, 2024A&A...689A.323M}.

Figure \ref{abundances_comp} compares the determined relative abundances of the main elements found in our two achondritic meteors with respect to the representative chondritic meteor (Table \ref{tab:abundances}), as well as the average elemental abundances in aubrite and eucrite meteorites relative to H chondrites from the literature. Consistent trends are observed in both datasets, further supporting our interpretation of the composition of these meteoroids. The most notable differences between the average abundances in meteorites and our achondritic meteors include lower abundances of Na in our achondritic meteors and a high Al/Mg abundance in our eucrite candidate. The determined Al abundances in our meteors are affected by the fact that strong Al lines were only observed in the eucrite candidate; they were faint or missing in the ordinary chondrite and aubrite candidate. The estimated Na abundances have higher uncertainties due to the saturation of the Na I-1 lines and the fact that only a few other Na I multiplets were available for fitting.

\section{Physical properties} \label{sec:physical}

While our study implies that the unambiguous identification of most achondritic meteoroids requires capturing good-quality emission spectra, we next focused on analyzing the ablation and fragmentation properties of these meteors. We aimed to examine whether achondritic meteoroids exhibit different atmospheric behavior and if the inferred physical properties are consistent with the estimated composition type.

\begin{table*}[ht]
\centering
\caption{Trajectory parameters and physical properties of the studied meteoroids.}
\label{tab:physical}
\resizebox{.8\textwidth}{!}{%
\begin{tabular}{lllllllll}
\hline \\[-5pt]
Meteor code & Type & Mag. & Mass (kg) & $\rho$\,(g\,cm\textsuperscript{-3}) & v\textsubscript{i}\,(km\,s\textsuperscript{-1}) & H\textsubscript{B}\,(km) & H\textsubscript{E}\,(km) & P\textsubscript{E}\, (Type) \\ [2pt]
\hline \\ [-5pt]
M20231120\_114953 (ACH1) & Aubrite & -8.2 ± 1.0 & 17.4 ± 8.7 & - & 13.18 ± 0.20 & - & - &  - \\ [2pt]
M20241012\_113432 (ACH2) & Eucrite & -7.6 ± 0.2 & 0.4 ± 0.1 & 3.16 ± 0.10 & 27.25 ± 0.06 & 89.15 ± 0.08 & 33.76  ± 0.05 & -3.99 (I)\\ [2pt]
M20240807\_093858 & O. chondrite & -9.7 ± 0.4 & 0.8 ± 0.1 & 3.58 ± 0.10 & 27.31 ± 0.07 & 89.89 ± 0.15 & 33.46 ± 0.05 & -3.93 (I) \\ [1pt]
\hline
\end{tabular}}
\tablefoot{"Meteor code" encodes the observation time in the format {\small $MYYYYMMDD\_HHMMSS$}. "Type" denotes the inferred composition class. Mag. is the absolute magnitude, Mass the entry photometric mass, $\rho$ the inferred bulk density, $v_i$ the entry speed, $H_B$ and $H_E$ the beginning and terminal heights, and $P_E$ the material strength parameter and type following the classification of \citet{1976JGR....81.6257C}.}
\end{table*}

\begin{figure*}[t]
\centering
\makebox[\textwidth][c]{
  \begin{minipage}[b]{0.45\textwidth}
    \centering
    \includegraphics[width=\linewidth]{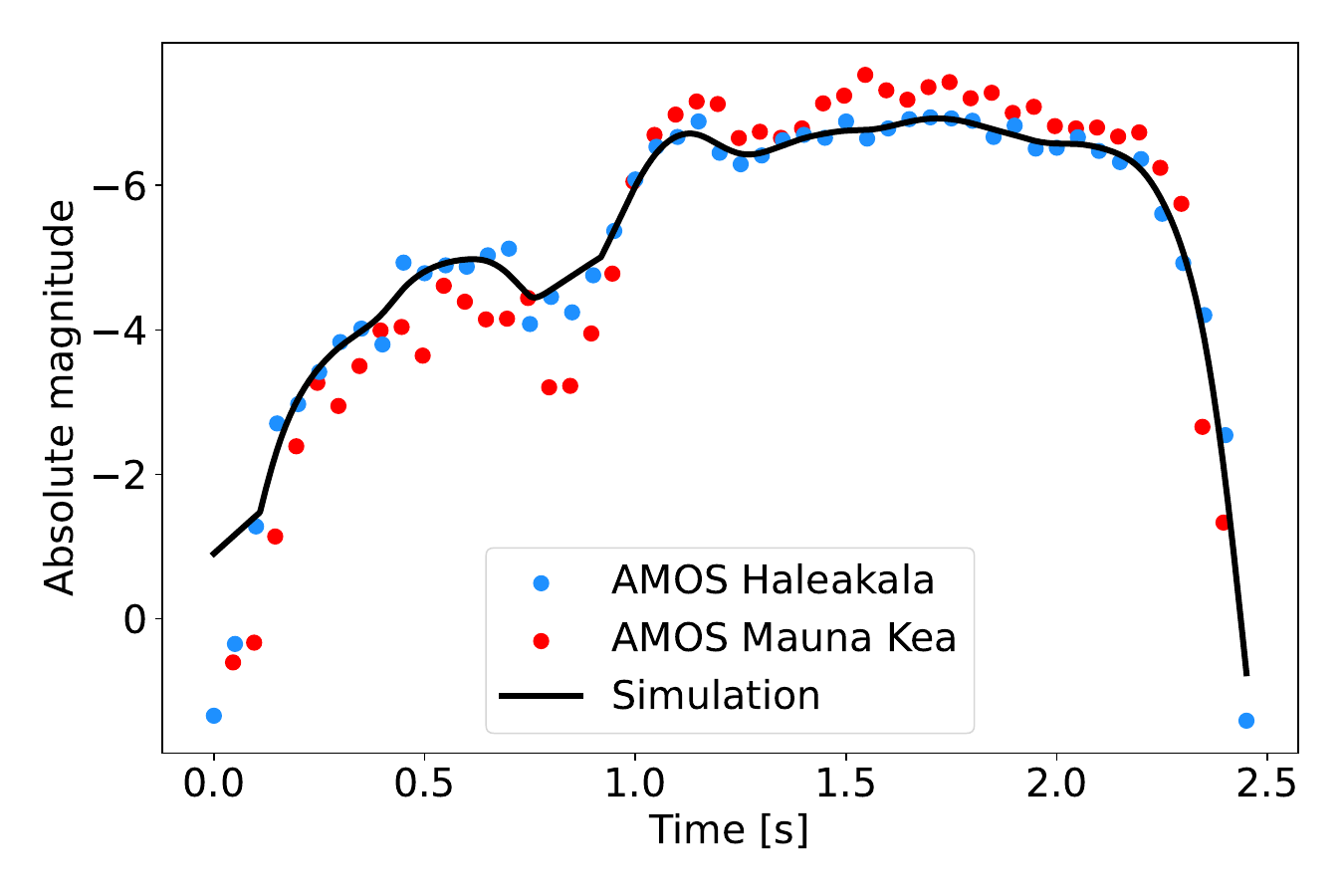}
  \end{minipage}
  \hspace{0.025\textwidth}
  \begin{minipage}[b]{0.45\textwidth}
    \centering
    \includegraphics[width=\linewidth]{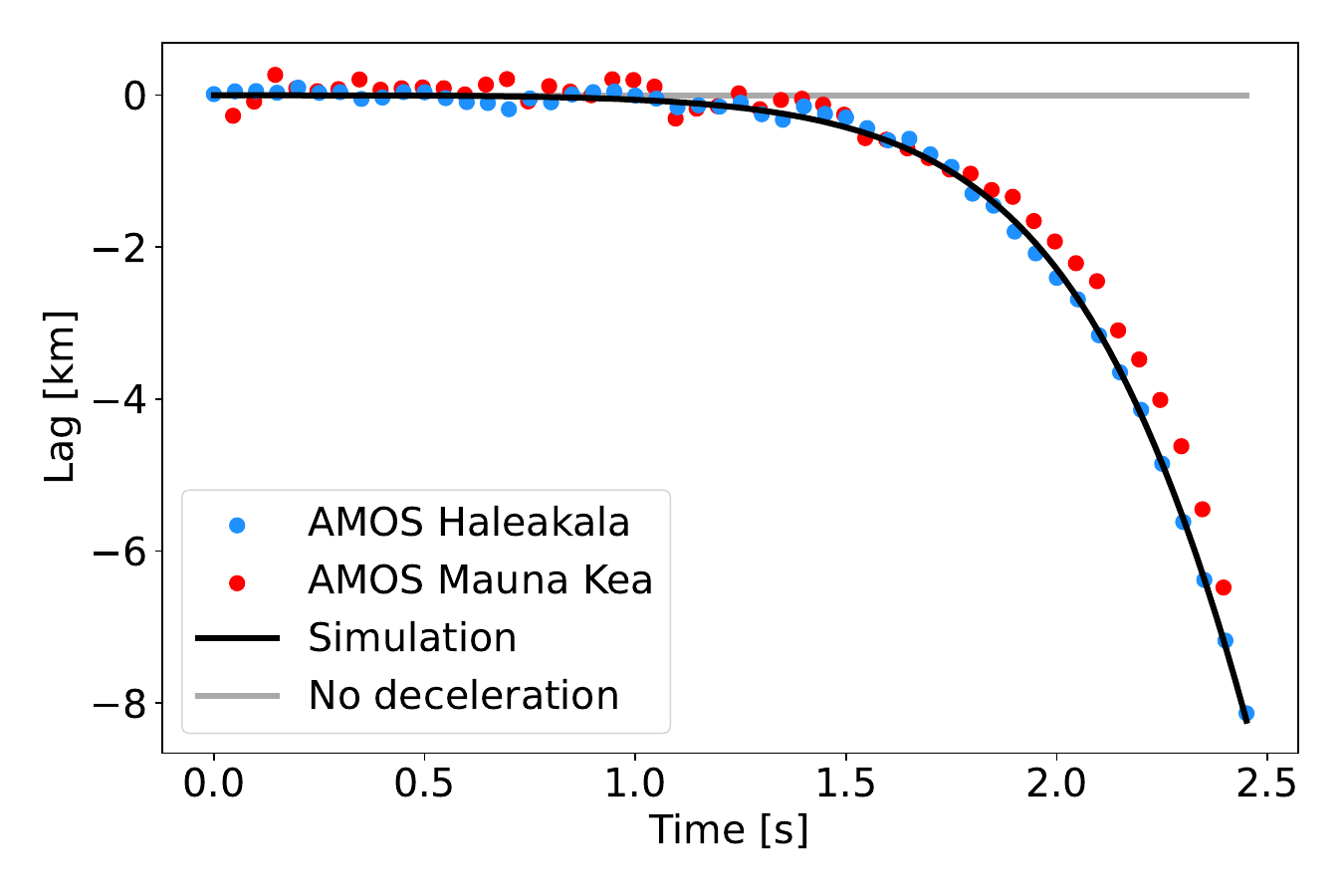}
  \end{minipage}
}
\caption{Final simultaneous fit of the light curve (left panel) and the deceleration profile (right panel) of the eucrite candidate based on the model of meteoroid erosion. The simulation is fitted to the Haleakala data, which are more precise because this station was closer to the meteor. Lag is defined as the difference between the distance traveled by the observed meteor and the distance traveled by a fictitious non-decelerating meteor.}
\label{eucrite_fit}
\end{figure*}

\begin{figure}[]
\centerline{\includegraphics[width=.95\columnwidth,angle=0] {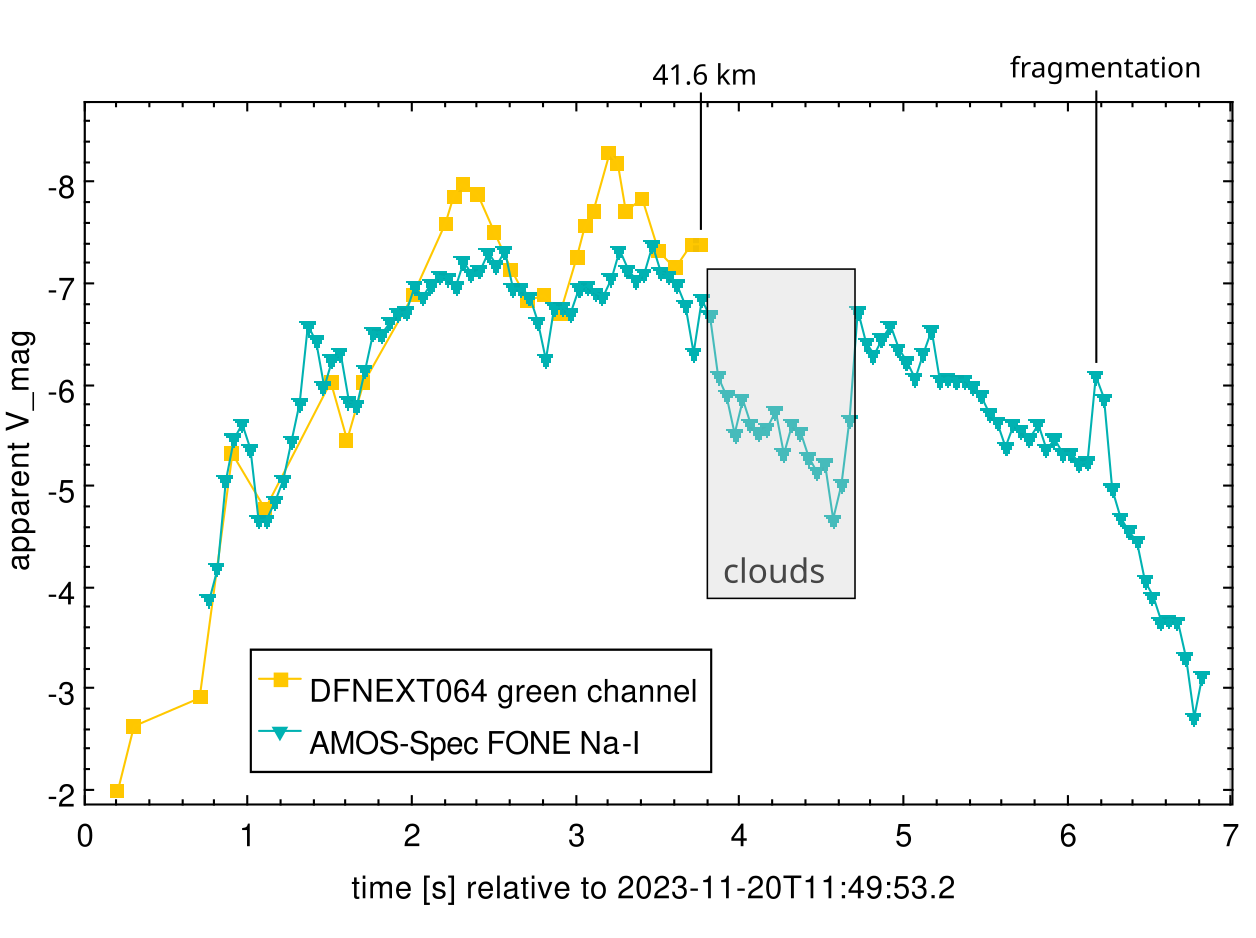}}
\caption[f1]{Apparent light curve of the ACH1 fireball (aubrite candidate) from Forrest Airport: the DFN long-exposure (green channel) and monochromatic light curve of Na I-1 emission line from the AMOS spectrograph (scaled to match the DFN-calibrated light curve).}
\label{aubrite_lightcurve}
\end{figure}

The light curve of the eucrite candidate (ACH2) already shows atypical features -- a dip in brightness of $\approx$\,1 magnitude occurring after $\approx$\,0.8\,s of the total 2.5\,s ablation interval (Fig. \ref{eucrite_fit}). This feature was observed by both AMOS stations in Hawaii and may indicate a transition phase in the ablation of different meteoroid components. The spectrum was only captured for the brighter second half of the meteor after the dip, so the direct confirmation of differential ablation from monochromatic light curves of individual chemical species is not possible. Nevertheless, a plausible interpretation of the $\approx$\,1 mag dip is a transition between the ablation of an initially more volatile fine-grained surface component and the onset of ablation of a more refractory interior. Once the outer layer was removed, the remaining material required additional heating before efficient mass loss resumed, producing a short-lived decrease in luminosity. An alternative explanation is a brief reduction in the effective ablation area after a shallow surface layer detached (spallation), which would have momentarily decreased the luminosity of the main body without producing a fragmentation flare.

This light curve shape proved challenging to fit with basic ablation models (single-body or gross fragmentation). We attempted to fit the light curve using a differential ablation model assuming two individual components defined by different ablation coefficients, densities, and masses. Even with this model, a satisfactory fit was only achieved when assuming two highly distinct components, which we did not consider physically justifiable. The best fit of the light curve and deceleration profile was finally achieved using a procedure based on the model of meteoroid erosion \citep{2007A&A...473..661B}. After a manual trial-and-error search for free parameters in the model, we produced the final fit of the light curve and lag profile shown in Fig. \ref{eucrite_fit}.

The results of our ablation and fragmentation modeling indicate that the meteoroid material was very compact, with only 14\% of its mass lost through erosion. We determined an ablation coefficient of $\sigma$ = 0.010 $\pm$ 0.001 kg/MJ, and, adopting a drag coefficient $\Gamma = 1$ and a spherical shape factor $A = 1.21$, we obtained a bulk density of $\rho$ = 3.16 $\pm$ 0.10 g\,cm\textsuperscript{-3}. This value is consistent with the average bulk densities of HED meteorites \citep{2011M&PS...46..311M} and is close to values characteristic of higher-density eucrites. The compact material of this meteoroid was also supported by the material strength parameter \citep{1976JGR....81.6257C} $P_E$ = -3.98 $\pm$ 0.11 (Type I) and the low terminal height of 33.96 $\pm$ 0.10 km. Our modeling suggests that $\approx$30\,g of material may have survived the atmospheric ablation and dropped into the Pacific Ocean near Hawaii. In comparison, the ordinary chondrite meteoroid of a comparable size and speed we analyzed (Table \ref{tab:physical}) was found to have a higher bulk density of $\rho$ = 3.58 $\pm$ 0.10 g\,cm\textsuperscript{-3}, consistent with the average densities of ordinary chondrite meteorites \citep{2006M&PS...41..331C}.

The photometry of the aubrite candidate meteor (ACH1) was partially obstructed by local clouds, making the determination of its physical properties difficult. The first half of the fireball trajectory was captured by the DFN cameras and used to determine the meteoroid entry speed and orbit. For the second part of the trajectory, only the emission spectrum was captured. The overall light curve was reconstructed by combining the DFN light curve and a monochromatic light curve of the Na I-1 line from the AMOS spectrograph (Fig. \ref{aubrite_lightcurve}). Compared to the relatively smooth light curve of the proposed eucrite (Fig. \ref{eucrite_fit}), the aubrite light curve suggests much more frequent fragmentation events indicated by short-term flares. Severe fragmentation was also reported during the atmospheric ablation of the aubrite Ribbeck fall \citep{2024A&A...686A..67S}. The brightness peak observed by DFN is not well fitted by the Na I-1 light curve due to the saturation of the spectral line. The dip in brightness observed in the middle of the light curve (Fig. \ref{aubrite_lightcurve}) is mostly caused by cloud coverage and does not reflect actual ablation behavior. 

Although almost seven seconds of meteor ablation were captured spectrally, the very beginning and second half of the luminous trajectory were not recorded (the end of the fireball happened between exposures in the DFN still imagery, and the zero-order image was outside of the FOV of the AMOS-Spec systems). Hence, no reliable terminal height is presented (Table \ref{tab:physical}). The entry mass, although determined with a higher uncertainty, is assumed to be greater than 10\,kg. During the observed phase, the meteor lost only $<$ 10\% of its initial speed. By extrapolating the observed end height and assuming no deceleration, the terminal height was estimated to $\approx$ 22 km. While the real terminal height was likely higher due to deceleration, it is still assumed that this fireball dropped meteorites in the Nullarbor Plain. 

The uncertainty of the terminal height and location of the fragmentation (Fig. \ref{aubrite_lightcurve}) means constraining a searchable strewn field is difficult. Furthermore, if the surviving meteorite is indeed an aubrite, its fusion crust will not be black like other meteorite types. This already posed a challenge for the identification of the Ribbeck meteorite \citep{2024A&A...686A..67S}, but it would be even harder in this context. Due to the presence of numerous light-colored rocks in the area, both traditional human searches and the DFN's drone-based searching solution \citep{2022ApJ...930L..25A} would struggle to identify the meteorite from the background rocks.

\section{Orbits and dynamical evolution} \label{sec:dynamical}

The orbital elements of the two achondrites and the ordinary chondrite used for comparison are presented in Table \ref{tab:orbital}. The proposed aubrite originated from a short-period, low-eccentricity, low-inclination orbit ($a$\,=\,1.081\,au, $e$\,=\,0.237, $i$\,=\,2.06\,$^\circ$). 

Aubrites are thought to originate from enstatite-rich E-type asteroids, which are believed to be most concentrated in the Hungaria region (1.8\,au\,<\,$a$\,<\,2\,au, 16\,$^\circ$\,<\,$i$\,<\,35\,$^\circ$), based on matching aubrite cosmic-ray exposure ages with the timescales of dynamical pathways from Hungaria to Earth-crossing orbits \citep{2014Icar..239..154C}. However, the orbital inclination of our aubrite candidate does not indicate such an origin. The Hungarias are also not a dominant supplier of the overall flux of near-Earth objects \citep{Granvik:2018, Nesvorny:2023}, and while dynamical pathways from the Hungaria region decrease a meteoroid's semimajor axis, fewer than half of the escaping objects would be expected to have orbital inclinations reduced to below 10 degrees \citep{Granvik:2016}.

Several known E-type asteroids have been found in the orbital region close to our aubrite candidate. The orbit of our case is positioned between the closest known E-type asteroids, (4660) Nereus \citep{2004P&SS...52..291B} and 2015 TC25—a potential boulder from (44) Nysa and one of the smallest characterized near-Earth asteroids \citep{2016AJ....152..162R}. Other known Apollo-type E types with typically higher eccentricities and inclinations ($\geq$ 7~$^{\circ}$)  include 1998 WT24, (3103) Eger, and 2024 BX1, the recent impactor that brought the Ribbeck meteorite to Earth. The determined orbit in our case is consistent with the view that aubrite-producing material is not necessarily restricted to the Hungaria region but can originate directly from E-type asteroids on near-Earth orbits.

The orbital parameters of the proposed eucrite (Table \ref{tab:orbital}) place it in a region consistent with the $\nu_6$ secular resonance at the inner-belt edge ($a$\,=\,2.097\,au, $e$\,=\,0.784, $i$\,=\,3.69\,$^\circ$). The $\nu_6$ resonance is a common delivery mechanism of HED meteorites to Earth \citep{1997M&PS...32..903M} and is the expected source region of the currently known HED meteorites with orbits \citep{JenniskensDevillepoix:2025}. 

There are several confirmed V-type near-Earth asteroids in this orbital region (such as 3551 Verenia, 3908 Nyx, and 1980 WF), though they typically have lower eccentricities. Cosmic-ray exposure age analysis of a recovered meteorite from this event could help us determine whether the meteoroid escaped from the main belt via the $\nu_6$ resonance or originated from a V-type asteroid already in near-Earth space. However, any recovery from this fireball is effectively precluded by the inferred fall area in the Pacific Ocean.

\begin{table*}[ht]
\centering
\caption{Orbital elements and the Tisserand parameter with respect to Jupiter ($T_J$) of the studied meteoroids.}
\label{tab:orbital}
\resizebox{.98\textwidth}{!}{%
\begin{tabular}{llcccccccc}
\hline \\[-5pt]
Meteor code & Type & a (au) & e & q (au) & Q (au) & i ($^\circ$) & $\Omega$ ($^\circ$) & $\omega$ ($^\circ$) & T\textsubscript{J} \\ [2pt]
\hline \\ [-5pt]
M20231120\_114953 (ACH1) & Aubrite      & 1.081 ± 0.020 & 0.237 ± 0.015 & 0.825 ± 0.002 & 1.337 ± 0.037 & 2.060 ± 0.020 & 237.510 & 262.32 ± 1.43 & 5.70 ± 0.07 \\ [2pt]
M20241012\_113432 (ACH2) & Eucrite      & 2.097 ± 0.017 & 0.784 ± 0.002 & 0.452 ± 0.001 & 3.741 ± 0.035 & 3.693 ± 0.014 & 19.313 & 104.11 ± 0.12 & 3.27 ± 0.02 \\ [2pt]
M20240807\_093858 & O. chondrite & 1.685 ± 0.019 & 0.742 ± 0.002 & 0.434 ± 0.001 & 2.935 ± 0.036 & 13.449 ± 0.048 & 135.056 & 290.01 ± 0.28 & 3.83 ± 0.03 \\ [1pt]
\hline
\end{tabular}}
\end{table*}

\begin{figure*}[t]
\centering
\makebox[\textwidth][c]{
  \begin{minipage}[b]{0.32\textwidth}
    \centering
    \includegraphics[width=\linewidth]{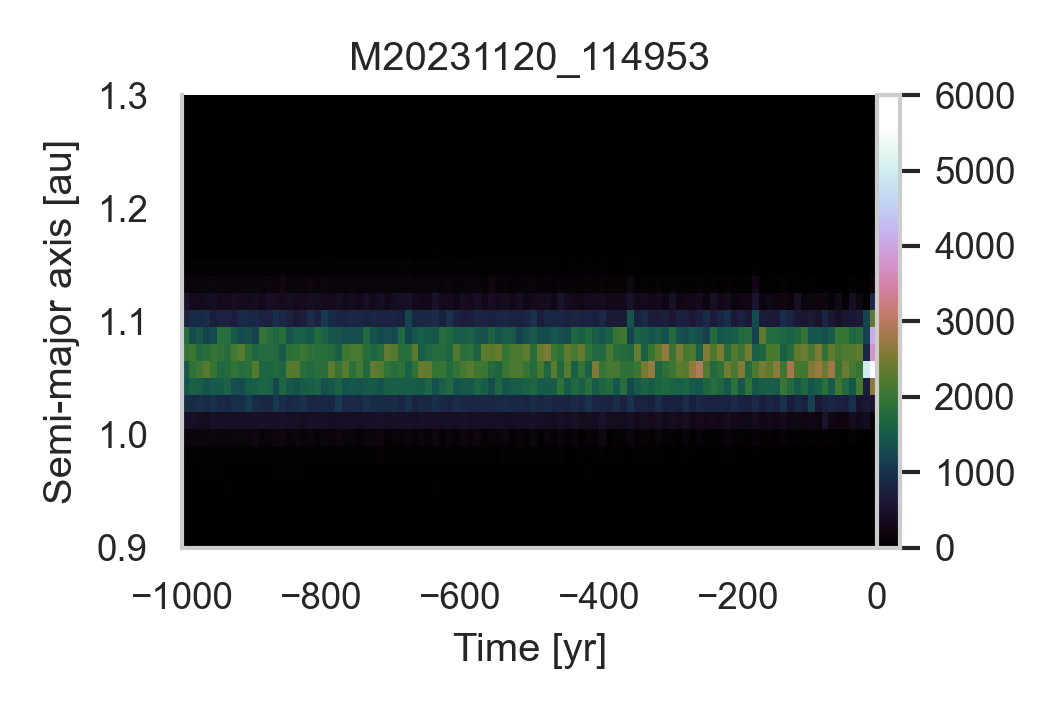}
  \end{minipage}
  
  \begin{minipage}[b]{0.32\textwidth}
    \centering
    \includegraphics[width=\linewidth]{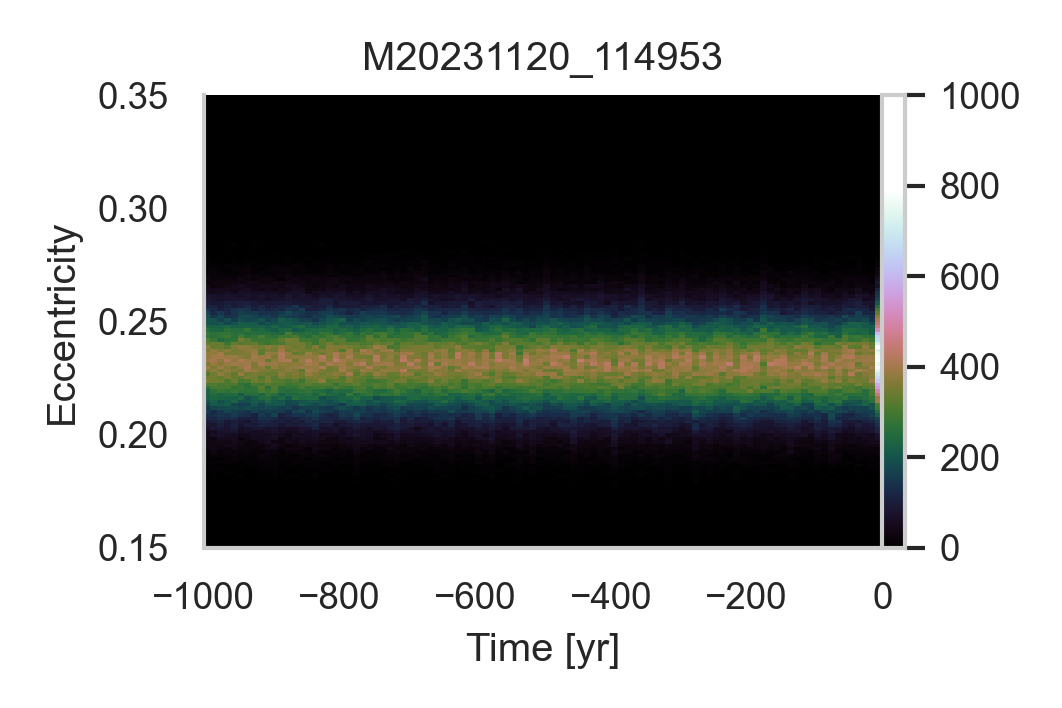}
  \end{minipage}
 
  \begin{minipage}[b]{0.32\textwidth}
    \centering
    \includegraphics[width=\linewidth]{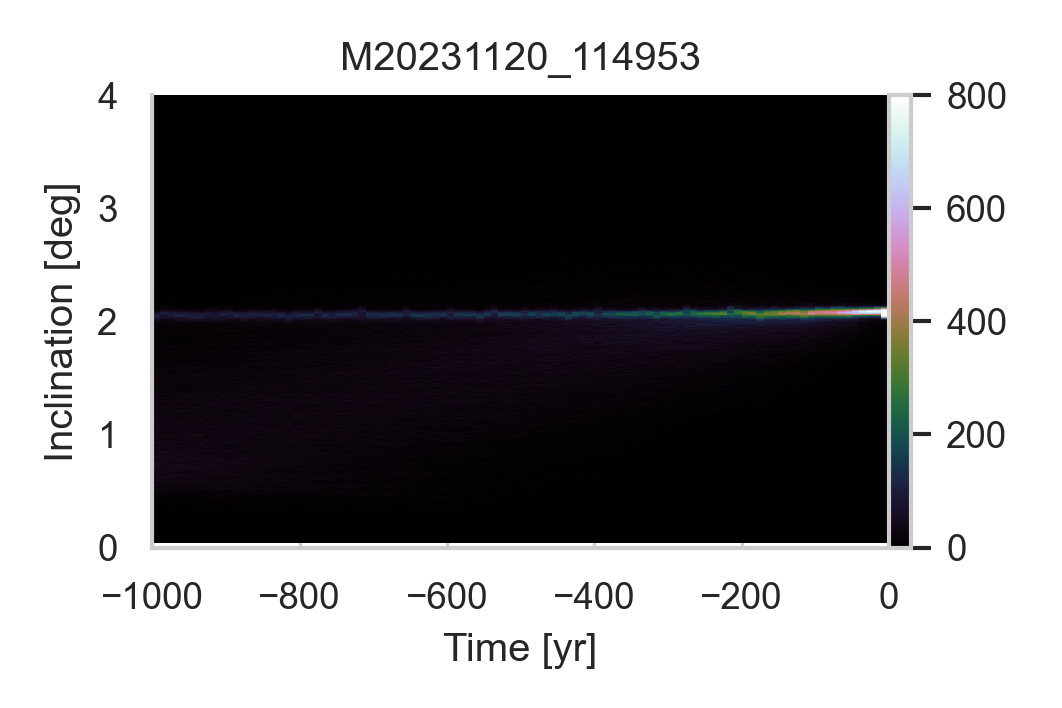}
  \end{minipage}
}

\makebox[\textwidth][c]{
  \begin{minipage}[b]{0.32\textwidth}
    \centering
    \includegraphics[width=\linewidth]{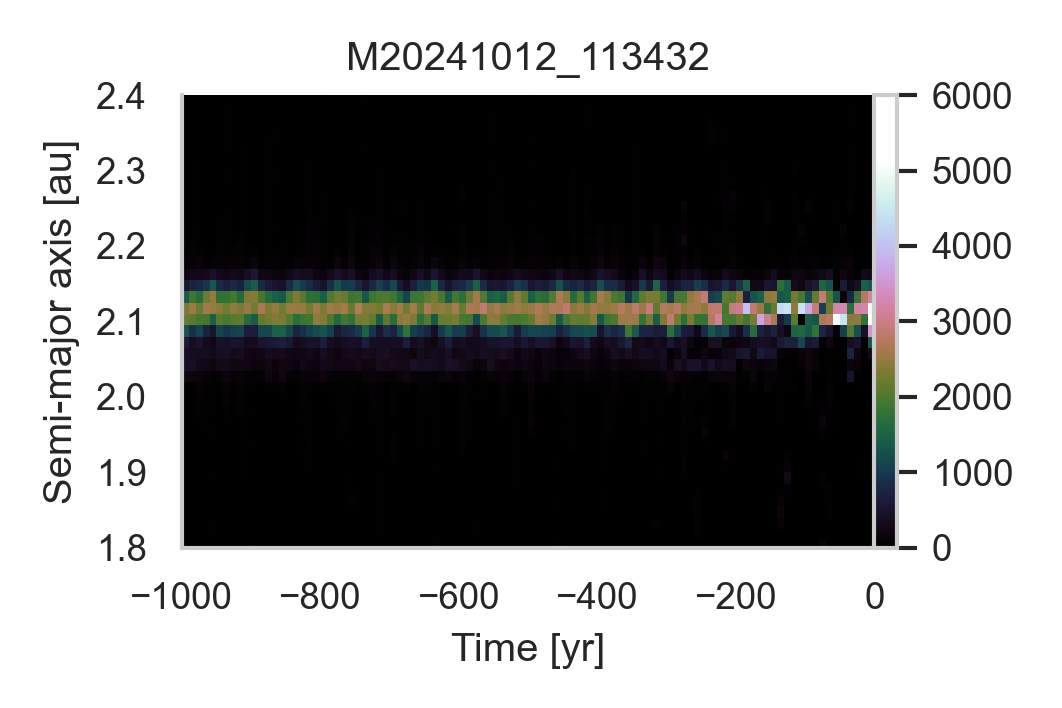}
  \end{minipage}
  
  \begin{minipage}[b]{0.32\textwidth}
    \centering
    \includegraphics[width=\linewidth]{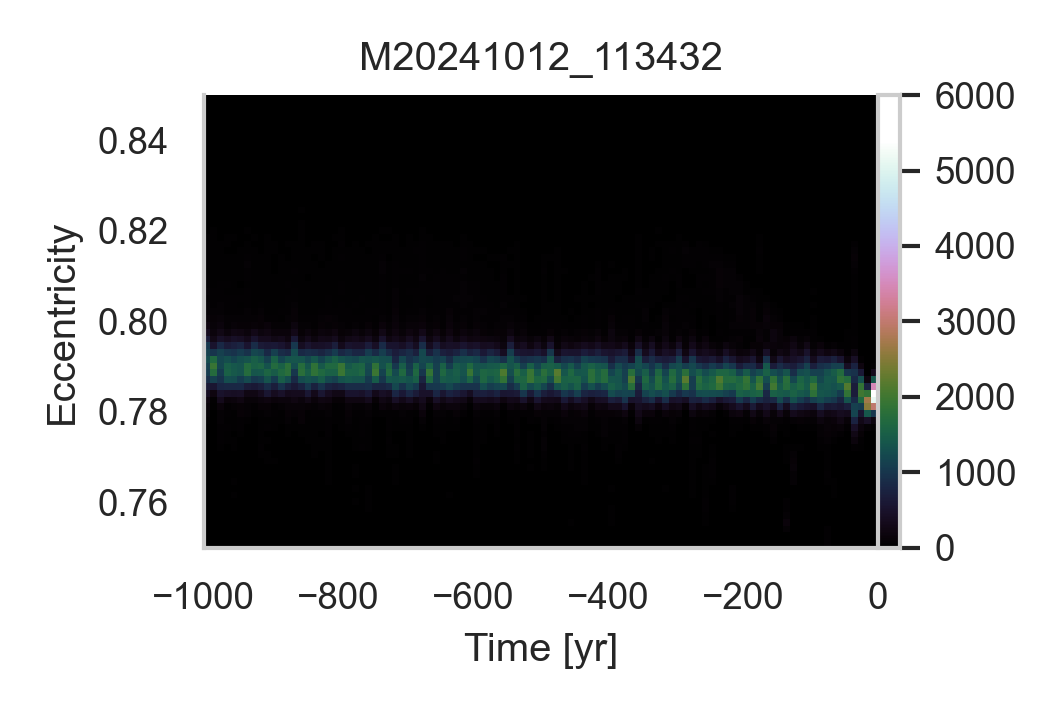}
  \end{minipage}
 
  \begin{minipage}[b]{0.32\textwidth}
    \centering
    \includegraphics[width=\linewidth]{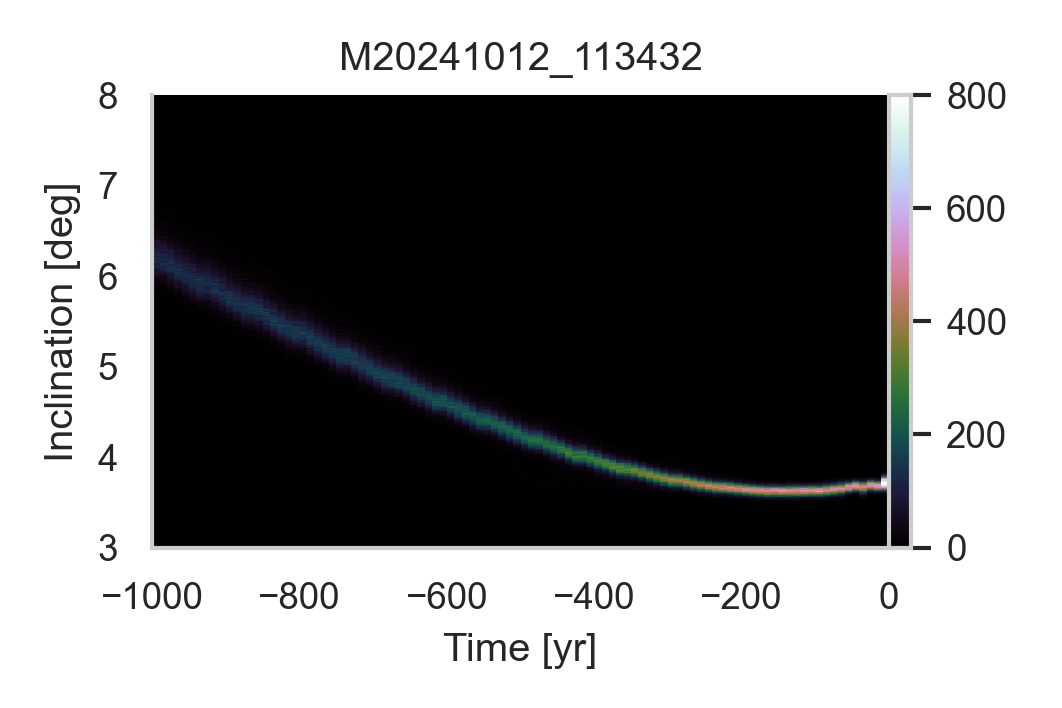}
  \end{minipage}
}

\caption{Dynamical evolution of the proposed aubrite ({\small M20231120\_114953}) and eucrite ({\small M20241012\_113432}) orbits based on backward integrations using 10000 simulated clones of their determined orbit (Table \ref{tab:orbital}) with readouts every ten years.}
\label{dynamical_evolution}
\end{figure*}

To assess the dynamical stability and origin of these meteoroids, we performed backward integrations of their orbits for 1000 years, accounting for the gravitational effects of all planets and large asteroids. The chosen integration duration was selected to be approximately ten times the Lyapunov timescales calculated for these orbits. We simulated 10000 clones with readouts every ten years. The initial clone distribution sampled the formal uncertainties of the measured orbital elements; nongravitational forces were not modeled. 

The massless clone particles in the simulations experienced numerous close encounters with massive bodies. Approximately 40\,\% of the eucrite particles and more than 80\,\% of the aubrite clones underwent at least one close approach with a planet. For the aubrite clones, the primary perturbing body was Earth, supplemented by a few encounters with Venus and Mars. In contrast, the evolution of the eucrite clones was influenced almost equally by Venus and Earth, with several additional encounters with Mars.

The evolution of orbital elements from our simulations is depicted in heatmaps in Fig. \ref{dynamical_evolution}. The short-term evolution of the eucrite orbit indicates a steady decrease in inclination. The semimajor axis and eccentricity are relatively stable, though some dispersion of the clones around the nominal orbit is observed. The increase in eccentricity potentially caused by the $\nu_6$ resonance is not evident on this short timescale. The diffusion of clone orbits is also apparent for the aubrite candidate (Fig. \ref{dynamical_evolution}), though most clones remain around the nominal orbit. This implies a mild dynamical instability for the orbit over $10^3$-year timescales, particularly in inclination, but does not contradict a recent delivery from the indicated source regions.

\section{Conclusions} \label{sec:conclusions}

The diversity and abundance of individual meteoroid composition types in the interplanetary space, partly reflecting all meteorite classes found on Earth, are not yet well constrained. This study presents a first step in our efforts to utilize multidisciplinary meteor observations for the classification of meteoroid composition types. From an analysis of 180 high-quality meteor spectra, the first two achondrites in our database were identified. Both were likely meteorite droppers. 

The spectral analysis revealed atypical diagnostic emission features for the two achondritic meteoroids, particularly the line intensities of Mg, Fe, Si, Ca, Al, Ti, and Li, which indicate compositions consistent with aubritic and eucritic materials. The relative elemental abundances measured using a radiative-transfer spectrum model support this classification. An unexpected enhancement in Ca, Mn, and Ti was observed in the aubrite-like spectrum, with time-resolved emission behavior distinct from the proposed eucrite spectrum. We interpret this as the rapid volatilization of highly reduced Ca- and Mn-bearing sulfide inclusions (e.g., oldhamite) rather than a bulk enrichment of these elements. This indicates that transient spectral spikes in refractory elements can trace internal heterogeneity within achondritic meteoroids, beyond what is inferred from their average composition.

The aubrite candidate originated from a short-period, low-eccentricity, low-inclination orbit similar to some known E-type near-Earth asteroids. While the atmospheric ablation data were limited, the partially recorded light curve shows signs of significant fragmentation, similar to the Ribbeck aubrite fall \citep{2024A&A...686A..67S}. The original mass of the impactor was over 10\,kg. Potential meteorites were not yet recovered due the uncertainty in terminal height. 

The eucrite candidate originated from an orbit affected by the $\nu_6$ resonance in the inner main belt, a common delivery mechanism of HED meteorites, and exhibited ablation behavior corresponding to a compact material with low erosion and an estimated bulk density of $\approx$ 3.16 $\pm$ 0.10 g\,cm\textsuperscript{-3}. Meteorites of a few tens of grams likely dropped into the Pacific Ocean near Hawaii.

Both achondritic meteoroids exhibited atypical light curves -- a dip in brightness separating two phases in the eucrite-candidate meteor and frequent short fragmentation flares in the aubrite candidate. Additional events with similar compositions are needed to determine whether these features represent the characteristic ablation behavior of these achondrite classes or are case-specific. While the observed ablation and fragmentation behavior, together with the inferred bulk densities and material strength parameters, is consistent with the estimated compositions, our results indicate that the reliable identification of achondrites from meteor observations will generally require analysis of their emission spectra.

Apart from the recent Ribbeck meteorite fall, the presented cases are the first achondrite meteors to have been analyzed in detail. Further observations, ideally validated by recovered meteorite falls, will help constrain the diversity of atypical meteoroid types and their behavior during atmospheric ablation. Such classifications, especially when paired with the orbital context, can directly inform material source-region links and recovery prioritization within fireball networks.

\section*{Data availability}

The datasets generated as part of this study are available from the corresponding author upon reasonable request. 

\begin{acknowledgements}

This work was supported by ESA under contract no. 4000140012/22/NL/SC/rp, by the Slovak Grant Agency for Science grant VEGA 1/0218/22, and by the Slovak Research and Development Agency grants VV-MVP-24-0232 and APVV-23-0323. M. Paprskárová and F. Hlobik were supported by the Comenius University Grants no. UK/1284/2025 and UK/1346/2025, respectively.
S. E. Deam acknowledges the support of an Australian Government Research Training Program Scholarship. A. Pisarčíková conducts her research under the Marie Skłodowska-Curie Actions - COFUND project, which is cofunded by the European Union (MERIT - Grant Agreement No. 101081195).
     
\end{acknowledgements}

\bibliographystyle{aa}
\bibliography{references}

\end{document}